\documentclass{aa}
\pdfoutput=1
\usepackage{color}
\usepackage{graphicx}
\usepackage[dvipsnames]{xcolor}
\usepackage{mathrsfs}
\usepackage{esvect}
\usepackage{multirow}
\usepackage{array}
\usepackage[normalem]{ulem}
\usepackage{url}
\usepackage[amssymb]{SIunits}
\usepackage{subcaption}
\usepackage{txfonts}

\usepackage{aas_macros}
\usepackage{natbib,twoopt}

\makeatletter
\renewcommand*\aa@pageof{, page \thepage{} of \pageref*{LastPage}}
\makeatother
\usepackage[colorlinks=true, linkcolor=blue, citecolor=NavyBlue, urlcolor=NavyBlue]{hyperref}

\usepackage{etoolbox}
\makeatletter
\patchcmd\@combinedblfloats{\box\@outputbox}{\unvbox\@outputbox}{}{%
   \errmessage{\noexpand\@combinedblfloats could not be patched}%
}%
 \makeatother
 
\newcommand{\para}[1]{\paragraph{\textbf{#1}}}

\newcommand{\msun}{\mbox{\rm M$_\sun$}}
\newcommand{\rsun}{\mbox{\rm R$_\sun$}}
\newcommand{\lsun}{\mbox{\rm L$_\sun$}}
\newcommand{\sub}{_\mathrm}
\newcommand{\kms}{\kilo\metre\usk\reciprocal\second}
\newcommand{\vsini}{v\,\sin\,i}
\newcommand{\Req}{R_{\mathrm{eq}}}
\newcommand{\Rpol}{R_{\mathrm{pole}}}
\newcommand{\Teq}{T_{\mathrm{eq}}}
\newcommand{\Tpol}{T_{\mathrm{pole}}}

\newcommand{\gpol}{g_{\mathrm{pole}}}
\newcommand{\veq}{v_{\mathrm{eq}}}
\newcommand{\Xc}{X_{\mathrm{c}}}

\newcommand{\OmegaK}{\Omega_{\mathrm{K}}}
\newcommand{\Teff}{T_{\mathrm{eff}}}
\newcommand{\geff}{g_{\mathrm{eff}}}
\newcommand{\logg}{\log g}
\newcommand{\Rc}{R_{\mathrm{c}}}

\newcommand{\rt}{r_{\theta}}
\newcommand{\rtt}{r_{\theta\theta}}

\newcommand{\sint}{\sin\theta}
\newcommand{\cost}{\cos\theta}

\bibpunct{(}{)}{;}{a}{}{,} 
\makeatletter
\newcommandtwoopt{\citeads}[3][][]{\href{http://adsabs.harvard.edu/abs/#3}%
{\def\hyper@linkstart##1##2{}%
\let\hyper@linkend\@empty\citealp[#1][#2]{#3}}}
\newcommandtwoopt{\citepads}[3][][]{\href{http://adsabs.harvard.edu/abs/#3}%
{\def\hyper@linkstart##1##2{}%
\let\hyper@linkend\@empty\citep[#1][#2]{#3}}}
\newcommandtwoopt{\citetads}[3][][]{\href{http://adsabs.harvard.edu/abs/#3}%
{\def\hyper@linkstart##1##2{}%
\let\hyper@linkend\@empty\citet[#1][#2]{#3}}}
\newcommandtwoopt{\citeyearads}[3][][]{\href{http://adsabs.harvard.edu/abs/#3}
{\def\hyper@linkstart##1##2{}%
\let\hyper@linkend\@empty\citeyear[#1][#2]{#3}}}
\makeatother

\begin{document}

   \title{A realistic two-dimensional model of Altair\thanks{Based on VLTI observations performed at ESO, Chile under programme IDs 60.A-9164(A), 87.D-0150(A), and 094.C-0232(A).}}


   \author{K. Bouchaud\inst{1}\fnmsep\inst{2}
          \and
          A. Domiciano de Souza\inst{1}
          \and
          M. Rieutord\inst{3}\fnmsep\inst{4}
          \and
          D. R. Reese\inst{2}
          \and
          P. Kervella\inst{2}
          }

   \institute{Université Côte d'Azur, Observatoire de la Côte d'Azur, CNRS, Laboratoire Lagrange,
   			  Bd de l'Observatoire, CS 34229, 06304 Nice cedex 4, France
         \and
         	 LESIA, Observatoire de Paris, Université PSL, CNRS, Sorbonne Université, Univ. Paris Diderot, 				 Sorbonne Paris Cité, 5 place Jules Janssen, 92195 Meudon, France
         \and
             Université de Toulouse, UPS-OMP, IRAP, 31028 Toulouse,
             France
         \and
         	 CNRS, IRAP, 14 avenue Édouard Belin, 31400 Toulouse, France
             }

   \date{Received \today; accepted }

 
  \abstract
   {Fast rotation is responsible for important changes in the structure and evolution of stars and the way we see them. Optical long baseline interferometry (OLBI) now allows for the study of its effects on the stellar surface, mainly gravity darkening and flattening.}
   {We aim to determine the fundamental parameters of the fast-rotating star Altair, in particular its evolutionary stage (represented here by the core hydrogen mass fraction $\Xc$), mass, and differential rotation, using state-of-the-art stellar interior and atmosphere models together with interferometric (ESO-VLTI), spectroscopic, and asteroseismic observations.}
   {We use ESTER two-dimensional stellar models to produce the relevant surface parameters needed to create intensity maps from atmosphere models. Interferometric and spectroscopic observables are computed from these intensity maps 
   and several stellar parameters are then adjusted using the publicly available MCMC algorithm Emcee.
   }
   {We determined Altair's equatorial radius to be $\Req = 2.008 \pm 0.006\,$\rsun, the position angle 
   $PA = 301.1 \pm 0.3\degr$, the inclination $i = 50.7\pm 1.2\degr$, and the equatorial angular velocity $\Omega = 0.74\pm 0.01$ times the Keplerian angular velocity at equator. This angular velocity leads to a flattening of $\varepsilon=0.220\pm0.003$. We also deduce from the spectroscopically derived $v\sin i\simeq 243$~\kms, a true equatorial velocity of $\sim314$~\kms\ corresponding to a rotation period of 7h46m ($\sim$3 cycles/day). The data also impose a strong correlation between mass, metallicity, hydrogen abundance, and core evolution. Thanks to asteroseismic data, and provided our frequencies identification is correct, we constrain the mass of Altair to 1.86$\pm$0.03~\msun\ and further deduce its metallicity $Z=0.019$ and its core hydrogen mass fraction $\Xc=0.71$, assuming an initial solar hydrogen mass fraction $X=0.739$. These values suggest that Altair is a young star $\sim$100~Myrs old. Finally, the 2D ESTER model also gives the internal differential rotation of Altair, showing that its core rotates approximately 50\% faster than the envelope, while the surface differential rotation does not exceed 6\%.
   }
   {}


   \maketitle
%

\section{Introduction} \label{section:Introduction}
 A large fraction of intermediate-mass and massive stars have high rotation rates \citepads{2012A&A...537A.120Z,2013A&A...560A..29R}. Understanding the physics of their interiors requires an understanding of how rotation affects them. Indeed, at high initial masses, stars rotate (on average) rapidly, up to several hundred \kms\ for the fastest \citepads{2009LNP...765..207R}. Such high rotation rates, which sometimes bring stars close to break-up, have a strong impact on their internal structure and evolution. For example, unless they possess a strong magnetic field, and early-type stars rarely do, many of these stars exhibit signs of differential rotation \citepads{2007AN....328.1034R}. This differential rotation induces meridional circulation and small-scale turbulence that carry matter and angular momentum \citepads{1953MNRAS.113..716M}. This effect is commonly called `rotational mixing', and while it brings fresh hydrogen to the core, increasing the time the star spends in the main sequence (MS) phase (some stars may even skip the Blue Loop in the giant phase \citepads{2013A&A...553A..24G}), it also brings elements formed in the core of the star up to the surface, where they can be observed.
Abnormal abundances are thus expected in fast rotating stars, compared to the slowly rotating ones \citepads[e.g.][]{2000ApJ...544.1016H, 2000A&A...361..101M}. Yet, some fast rotators do not show the expected abnormal abundances, while some slow rotators do show abnormal abundances, as summarised in \citetads{2017A&A...603A..56C}. Obviously, rotational mixing still requires further investigation.


These examples show that taking rotation into account in stellar interior models is
paramount to a more accurate description of stellar evolution. Unfortunately,
as rotation has two-dimensional (even three-dimensional) effects, it cannot be easily
accounted for by 1D models. For instance, the 1D MESA models (e.g. \citeads{2018ApJS..234...34P}) include rotational mixing as a pure diffusive process while advection is known to be essential in the transport of angular momentum \citepads{1992A&A...265..115Z}. The difficulty is increased when one has to deal with data that are directly influenced by the non-spherical nature of rotating stars, like spectro-interferometric observables. For example, \citetads{2002A&A...393..345D} performed a detailed study on how physical models can be used to investigate and constrain some model parameters from the observed interferometric signatures of rotation, namely rotational flattening and gravity darkening (GD). From the observational side, many results were obtained from spectro-interferometry, and \citetads{2012A&ARv..20...51V} give a review of some of the different attempts made on several stars. Although important advances were achieved thanks to these and other theoretical and observational studies, they are based on simplified models, not taking into account the 2D internal structure of the rotating stars.


Recent progress in programming techniques and computer power has enabled the creation of fully two dimensional stellar models by the ESTER code \citepads{2016JCoPh.318..277R}. ESTER models indeed predict the differential rotation profile and the associated meridional circulation of an early-type star at a given stage of its MS evolution. The solution given by the code is presently the steady state solution of an isolated rotating star. Time evolution has not been implemented yet. However, by tuning the hydrogen mass fraction in the convective core, a good approximation of an evolved state on the MS can be computed. \citetads{2013A&A...552A..35E} have shown that the fundamental parameters of three nearby rapidly rotating stars obtained with interferometry could be reproduced fairly well. The three stars are \object{$\alpha$  Leo} ($M\simeq 4.15\msun$), \object{$\alpha$ Lyr} ($M\simeq 2.2\msun$) and \object{$\alpha$ Oph} ($M\simeq 2.4\msun$). However, the fundamental parameters of these stars have been derived from interferometric data using simple stellar models, namely Roche models with uniform rotation. Gravity darkening, a non-uniform flux distribution on the stellar surface as a result of rotation, was modelled using the modified von Zeipel's law $\Teff\propto\geff^{\beta}$, with the exponent $\beta$ being adjusted. Hence, several approximations of stellar structure are presently used to interpret the interferometric data. One might therefore wonder what would be the fundamental parameters derived from interferometric data if more realistic 2D models were used. This, in turn, would inform us on the validity domain of 1D models and their approximations.

To test this new way of modelling fast rotators, we chose to focus on the well studied A7V star Altair (\object{$\alpha$ Aquilae}, \object{HD187642}) for which numerous data sets are available. 

To determine its parameters, we first endowed ESTER models with appropriate stellar atmosphere models. We thus obtained spectra and intensity maps of Altair, the latter being constrained with interferometric data. The results gave new fundamental parameters for Altair based on the most up-to-date two-dimensional models. Interestingly, the new models enable an interpretation of the $\delta$-Scuti type oscillations of Altair detected by \citetads{2005ApJ...619.1072B}.

The paper is organised as follows: in Sect. \ref{section:Data}, we list the observational data that we use. In Sect. \ref{section:ESTER}, the ESTER code is briefly described. We present the preliminary models that match the previous determinations of Altair's parameters (Sect. \ref{section:Altair}). We then focus on the combination of atmospheric models with interior models (ESTER and $\omega$-model) and compute interferometric observables and spectra from the resultant intensity maps, to be compared with the observational data (Sect. \ref{section:Atmosphere}). The model-fitting method we used and the results obtained are presented in sections \ref{section:Model Fitting} and \ref{section:Results}. Finally, we discuss several points of interest (Sect. 8) and give some prospects for the future that this work offers.


\section{Data} \label{section:Data}
\subsection{Interferometry}
Two sets of near-IR interferometric data from two different VLTI beam-combiners were used to study Altair: 8 observations from PIONIER (H band; 1.65 \micro\meter) and 6 observations from GRAVITY (K band; 2.2 \micro\meter). The PIONIER observations (listed in Table \ref{table:PIONIER}) were obtained from the JMMC\footnote{Jean-Marie Mariotti Center} web service OiDB\footnote{Optical interferometry DataBase} and combine two different observing programmes. The data reduction and calibration process for both programmes were done using the \textit{PNDRS} pipeline \citepads{2011A&A...535A..67L}. The first part is made of five observations done over two nights in September 2011. Seven channels in the H band are available. The second part is composed of three observations done in one night, in October 2014, for the Exozodi survey (the details of the observation are given in \citeads{2014A&A...570A.127M}). They were conducted with a compact configuration (AT) and three wavelength channels in the H band. This second part of PIONIER data was used despite the three-year gap with the first one, as the compact configuration will only be sensitive to large scales, such as the overall shape of the star, which does not change significantly in such a short time for Altair.

The second set of data is made of two nights of GRAVITY observations. They were obtained during the first Science Verification Time (SVT) of the instrument (programme ID : 60.A-9164(A)). They were reduced using the GRAVITY pipeline \citepads{2014SPIE.9146E..2DL}, with the star HD188310 used as calibrator, through reduction recipes made available by the GRAVITY consortium in their python toolkit\footnote{available at \url{http://version-lesia.obspm.fr/repos/DRS_gravity/python_tools/}.}.
Six squared visibilities and four closure phases were obtained with each observation, for both the p and s polarisation directions. This yields two data sets at high resolution in the K band ($R\sim 4000$), and two at low resolution for the fringe tracker. As there was no significant difference between the data sets for the two polarisations, we averaged them. Only the science detector data was used in our analysis.

\begin{figure*}
\centering
  \includegraphics[width=\textwidth]{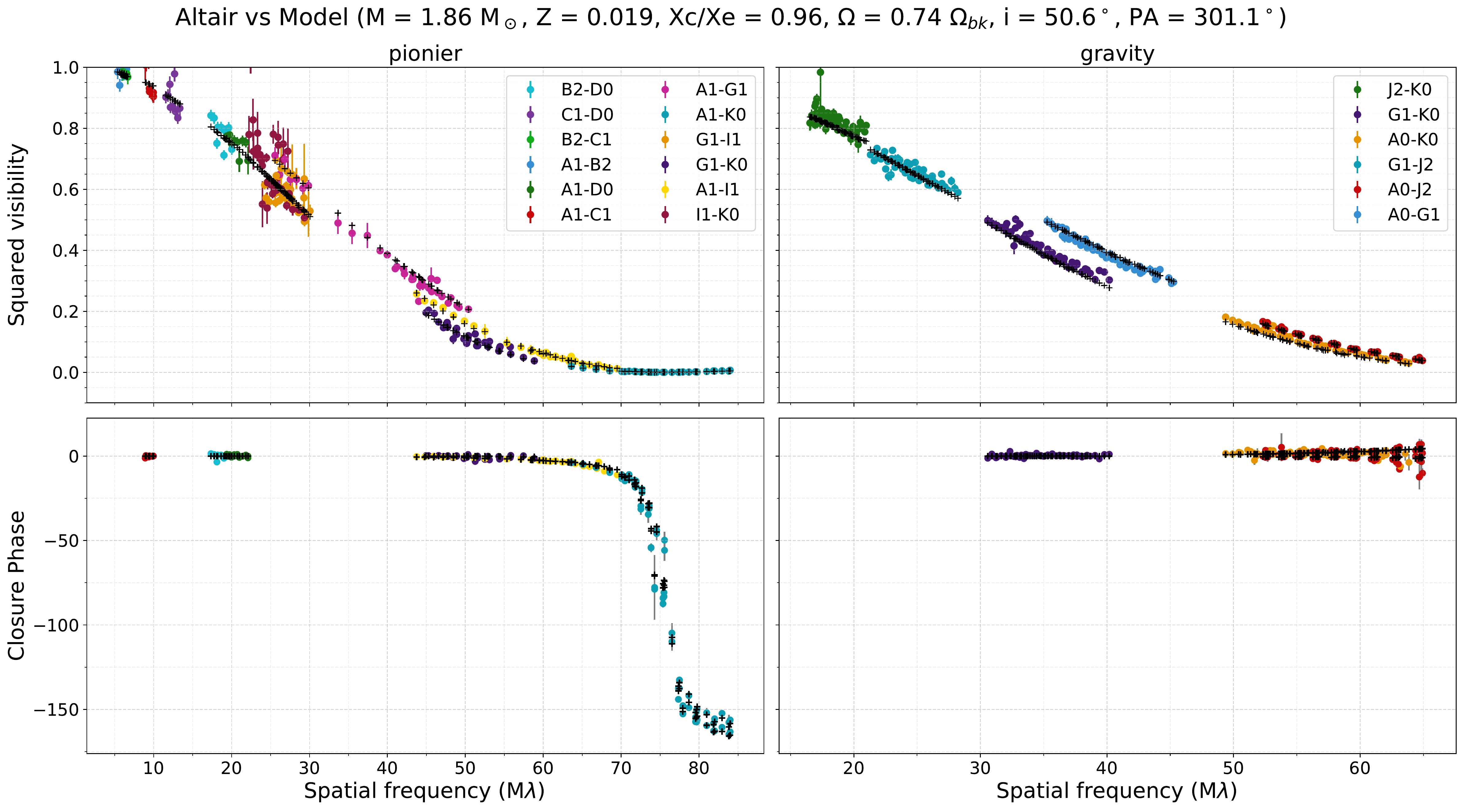}
	\caption{Interferometric observables (squared visibilities and closure phases) versus theoretical predictions from the best model (see Sect. \ref{section:Results}). Each colour indicates a different baseline, with the corresponding telescopes shown in legend.}
\label{image:fit_interfero}
\end{figure*}

\begin{figure*}[ht]
\centering
  \includegraphics[width=\textwidth]{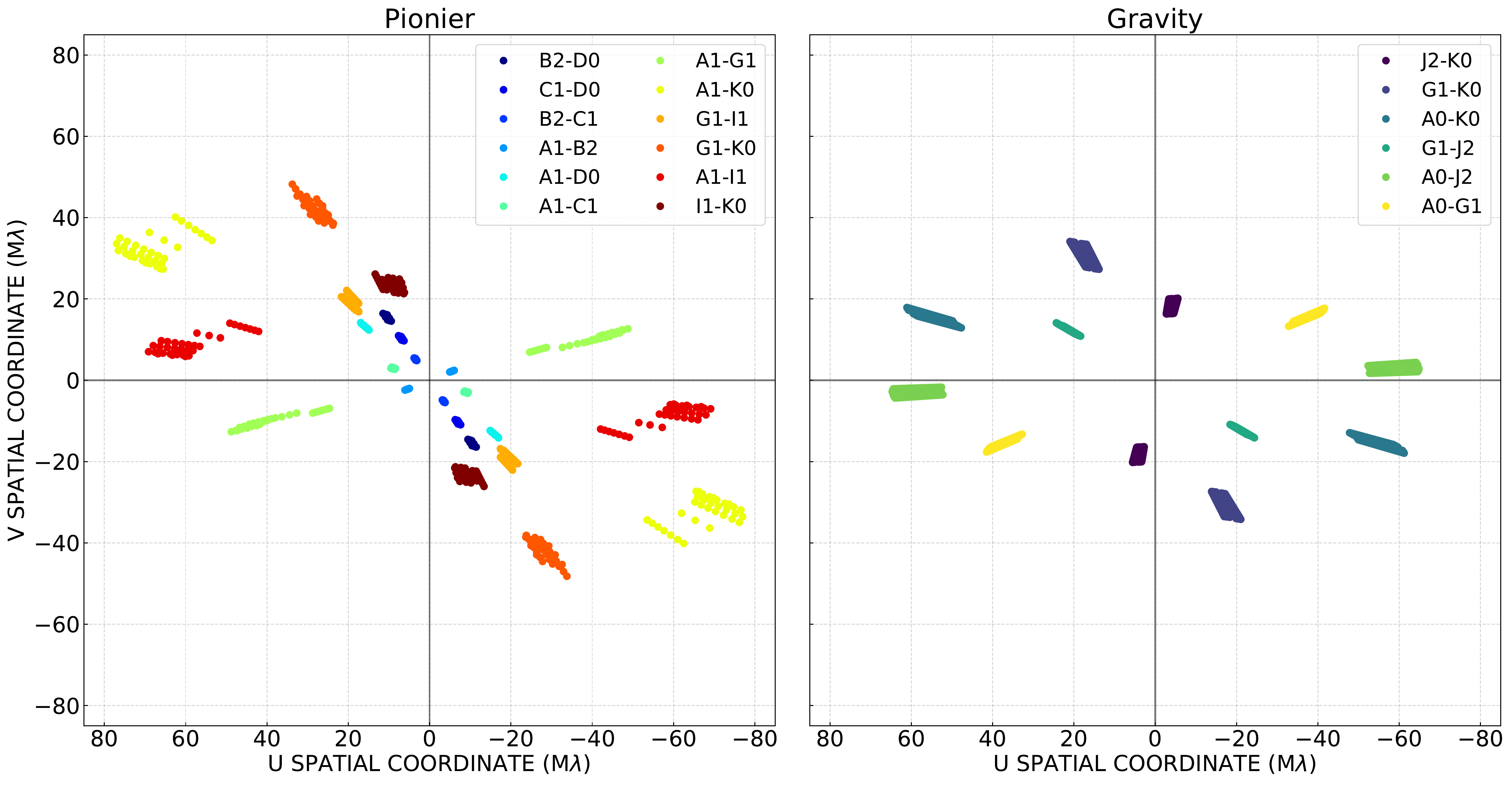}
	\caption{$uv$ coverage of PIONIER and GRAVITY observations of Altair.}
\label{image:uv_coverage_all}
\end{figure*}

The $uv$-plane for both instruments is shown in Fig. \ref{image:uv_coverage_all}, and the observables are shown in Fig. \ref{image:fit_interfero} along with the theoretical data.

\renewcommand{\arraystretch}{1.5}

\begin{table}
\centering
\caption{Altair's observations done with the PIONIER instrument at ESO's VLTI in the H band.}
\begin{tabular}{c c >{\centering}m{1.7cm} c} 
 \hline\hline
 Date  			  & Configuration & Ground baseline& PA ($\degr$)\\[0.5ex]
 \hline
 2011-09-23T00:43 &             &         & \\
 2011-09-23T01:49 &  			& 46.64m  & \multirow{2}*{See}\\
 2011-09-24T00:46 & A1-G1-I1-K0 & to      & \multirow{2}*{Fig. \ref{image:uv_coverage_all}}\\
 2011-09-24T01:18 & 			& 129.08m & \\
 2011-09-24T03:33 & 			&         & \\
 \hline
 2014-10-11T23:49 &             & 11.31m & \multirow{2}*{See}\\
 2014-10-12T00:12 & A1-B2-C1-D0 & to     & \multirow{2}*{Fig. \ref{image:uv_coverage_all}}\\
 2014-10-12T00:34 & 			& 36.09m & \\
 \hline
\end{tabular}
\label{table:PIONIER}
\end{table}

\begin{table}
\centering
\caption{Altair's observations done with the GRAVITY instrument at ESO's VLTI in the K band.}
\begin{tabular}{c c >{\centering}m{1.7cm} c} 
 \hline\hline
 Date  			  & Configuration 	           & Ground baseline        & PA ($\degr$)\\[0.5ex]
 \hline
 2016-06-16T06:50 &                            &                        & \\
 2016-06-16T06:58 &  						   & \multirow{2}*{48.86m}  & \\
 2016-06-16T07:06 & \multirow{2}*{A0-G1-J2-K0} & \multirow{2}*{to}      & See\\
 2016-06-18T06:06 & 					 	   & \multirow{2}*{129.34m} & Fig. \ref{image:uv_coverage_all}\\
 2016-06-18T06:14 & 					 	   &                        & \\
 2016-06-18T06:22 & 					       &                        & \\
 \hline
\end{tabular}
\label{table:GRAVITY}
\end{table}

\subsection{Spectroscopy}
\begin{figure}
\centering
  \includegraphics[width=0.48 \textwidth]{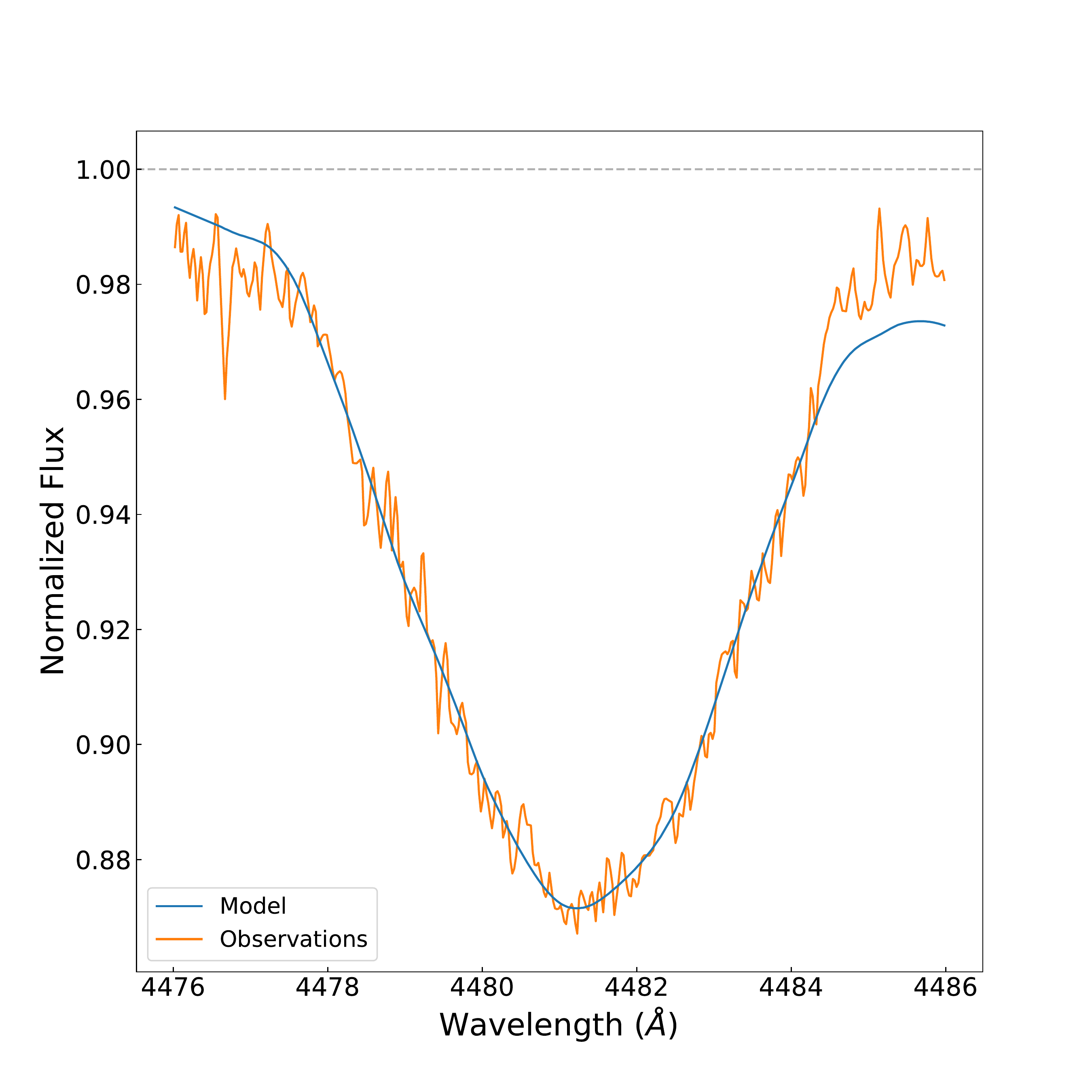}
	\caption{Observed and theoretical Mg\textsc{II} line at 4481 \angstrom\ of the best ESTER model (see Sect. \ref{section:Results}. The parameters of the corresponding model are shown in Table \ref{table:final_model}).}
\label{image:spectrum}
\end{figure}

The spectrum used for the spectroscopic analysis of the star was obtained on October 1st, 2003, using the ELODIE instrument at the Observatoire de Haute-Provence. After merging the orders of the initial echelle spectra, \citetads{2004A&A...428..199R} combined five single exposures, with a spectral resolution of 42000. They calibrated and normalised it using HD 118623 and $\gamma$ Boo as calibrators. The spectrum has a Signal-to-Noise Ratio (S/N) of 228 (average S/N for the five exposures), between $\lambda=3850$ and 6800 $\angstrom$.
As no error bars were computed in the process, we averaged the relative errors found in the files of the five single exposures, available in the `ELODIE archives'\footnote{\url{http://atlas.obs-hp.fr/elodie/fE.cgi?ob=objname,dataset,imanum&c=o&o=altair}}. 
The part of the spectrum used in our analysis is shown in Fig. \ref{image:spectrum} along with the theoretical line.

\section{ESTER : Modelling stellar interiors in 2D} \label{section:ESTER}




\subsection{Partial differential equations}

Several attempts at modelling stars in two dimensions have been made over the last 50 years
or so, but all have failed to describe the large-scale dynamics of the interior, with realistic
internal rotation. The aim of the ESTER project has therefore been to address this issue, thus leading to the eponymous code. We shall briefly describe the models here, but a more detailed presentation may be found in   \citetads{2013A&A...552A..35E} and \citetads{2016JCoPh.318..277R}.

ESTER models describe the (quasi) steady  state of an isolated, non-magnetic early-type star made of a convective core and a radiative envelope. The core is assumed to be isentropic, which is very close to what the mixing-length modelling says.

The four partial differential equations that determine such a model are :

\begin{equation}
\left\{ \begin{array}{l}
		\Delta\phi = 4\pi G\rho\\
    	\rho T \vec{v} \cdot \nabla S = - \nabla\cdot\vec{F} + \varepsilon_*\\
        \rho \vec{v}\cdot\nabla\vec{v} = - \nabla P - \rho\nabla \phi + \vec{F}_v\\
        \nabla\cdot(\rho \vec{v}) = 0
	\end{array}\right. 
    \label{eq:eq_ESTER}
\end{equation}

These are Poisson’s equation ($\phi$ is the gravitational potential, $\rho$ the density,
and $G$ the gravitational constant), the equation of entropy $S$ ($T$ is the temperature,
$\vec{v}$ the velocity, $\vec{F}$ the heat flux, and $\varepsilon_*$ the nuclear heat
sources), the momentum equation in an inertial frame ($P$ is the pressure and $\vec{F}_v$ the viscous force), and the equation of mass conservation.

These equations are completed by the expressions of the heat flux, the viscous
force, the energy generation, the equation of state and opacities. For these latter quantities OPAL tables are used \citepads{1996ApJ...456..902R}.

\subsection{Boundary conditions}
The system of equations needs to be completed with appropriate boundary conditions, in particular at the surface: (i) the gravitational potential must match that of a field in vacuum, vanishing at infinity; (ii) the velocity field must match stress-free conditions; and (iii) the temperature must meet the local black body radiation condition.

These conditions are applied on the bounding surface, chosen as the isobar where the
pressure is equal to the polar one, defined as
\begin{equation}
P_s = \tau_s\frac{\gpol}{\kappa_{\mathrm{pole}}},
\end{equation}
where $\tau_s$ is the optical depth that defines the polar photosphere, $g\sub{pole}$ is the polar gravity and $\kappa_{pole}$ is the polar opacity \citepads[for more details see][]{2013A&A...552A..35E}.
The surface radius, chosen as the radius of the polar isobar, coincides with the polar photospheric radius, but slightly differs from the actual photosphere radius elsewhere. However, it can be shown (see appendix \ref{appendix:surface_radius}) that as long as the flattening of the star, defined as
\begin{equation}
    \varepsilon = 1 - \frac{\Rpol}{\Req},
    \label{eq:epsilon}
\end{equation}
does not exceed 0.2, the difference between the two radii is less than one percent. Interestingly, $\varepsilon \sim 0.2$ is the value obtained for the models of Altair presented in Sect. \ref{section:Results}.


Finally, either the total angular momentum or the surface equatorial velocity $V_{\mathrm{eq}}$ has to be specified.

\subsection{Numerical solver}


For the numerical part, the star is subdivided into multiple domains in the radial direction.  In each domain, a Gauss-Lobatto collocation grid appropriate for spectral methods based on Chebyshev polynomials is used.  The advantage of a multi-domain approach is that even for spectral methods, it is possible to increase the resolution where it is needed, in particular near the stellar surface where rapid variations occur.  An appropriate set of coordinates, which adapts itself to the centrifugal distortion of the star, is used.

The set of equations (\ref{eq:eq_ESTER}) is then solved over the stellar interior using
Newton's method. This iterative method has a quadratic convergence if the initial
solution given in input is not too far from the real one. Usually, a 1-D model
computed with ESTER is given as input for the computation of a 2D one when the
rotation velocity does not exceed 50 to 70 \% of the break-up value (the break-up rotation velocity is defined as the velocity at which the centrifugal force and the gravitational force at the equator are equal). Above that rotation rate, an intermediate 2D model must be computed first and given as input.

\subsection{Limitations}
Apart from the already mentioned assumptions concerning the star, such as the absence
of a magnetic field, ESTER models face other limitations. The major one is that convection
in surface layers (other than the core) has not yet been implemented, thus stars with
a mass below about 1.6 \msun\ and evolved stars cannot be modelled, as convective layers start
to form in these stars just below the surface. Altair is thus a good test for ESTER since it is a well studied star and its mass is close to the lower mass limit manageable by the code. On the high mass side, the limit is around 40 \msun\ presently (e.g. \citeads{2019A&A...625A..88G}).

\section{Altair's fundamental parameters} \label{section:Altair}

Altair is a nearby (5.13 $\pm$ 0.02 pc, \citeads{2007A&A...474..653V}) rapidly rotating star that has already been studied extensively in OLBI. \citetads{2001ApJ...559.1155V} were the first ones to measure the oblateness of Altair, using data from the Palomar Testbed Interferometer (PTI). They found an oblateness $\varepsilon$ (Eq.~\ref{eq:epsilon}) of 0.122 $\pm$ 0.022, and also derived a projected rotation velocity $\vsini$, independently of spectroscopic analyses, of 210 $\pm$ 13 \kms. This value falls within the range of velocities determined through different spectral line studies, from 190 \kms\ \citepads{1984ApJ...286..741C} to 250 \kms\  \citepads{1968MNRAS.140..121S}, the latest estimate being 227 $\pm$ 11 \kms\  \citepads{2004A&A...428..199R}.

\citetads{2004ApJ...612..463O} then showed evidence of gravity darkening from NPOI data (Navy Prototype Optical Interferometer). \citetads{2005A&A...442..567D} confirmed that Altair was compatible with the von Zeipel relationship between $T\sub{eff}$ and $g$ for hot stars ($T\sub{eff} \propto g^{\beta}$, with $\beta=0.25$) after adding VLTI's VINCI data to the two previously mentioned data sets, analysing the star in different spectral bands. They also gave a broad estimate of the inclination angle $i$ (angle between the polar axis and the line of sight), between 40 and 65\degr. \citetads{2006ApJ...636.1087P} gave a more precise value of 63.9 $\pm$ 1.7\degr.

The latest interferometric study of Altair led to the first `direct' imaging (via image reconstruction techniques), by \citetads{2007Sci...317..342M}, exploiting data from the CHARA array's instrument MIRC. They found a departure from von Zeipel's theorem, with a value of $\beta=0.19$ in their models allowing a better fit to the data. 

Along with interferometry, other observation techniques give us additional information on Altair. In addition to the projected velocity, \citetads{2004A&A...428..199R} determined its inclination to be greater than 68\degr at a 1$\sigma$-level (45\degr at a 2$\sigma$-level). On the asteroseismic side, let us mention the works of \citetads{2005ApJ...619.1072B} and \citetads{2005A&A...438..633S} who showed with data from the WIRE satellite that Altair exhibits $\delta$ Scuti-type pulsations.\\


Using two-dimensional models from the ESTER code, we aim at further improving the determination of Altair's fundamental parameters by better modelling the effects of fast rotation. This can be done by comparing the observational data with our models, but it requires replacing the Roche models (simple and fast to create) with the more numerically demanding ESTER models. To test the feasibility of this processing we first derived an ESTER model of Altair by a manual adjustment of its parameters to the parameters found by \citetads{2007Sci...317..342M}. The result given in Table \ref{table:first estimate} shows that ESTER models can reproduce fairly well the parameters obtained by \citetads{2007Sci...317..342M}. We note that the equatorial velocity is slightly off the error bars and that the derived mass is smaller than the one derived by Monnier et al. Nevertheless, this first attempt encouraged us to take up the challenge of deriving again, ab initio, the fundamental parameters of Altair from the above data using ESTER models.


\begin{table}
\centering
\caption{Comparison between observationally derived parameters of Altair by \citetads{2007Sci...317..342M} and a manual fit to effective temperatures, radii and equatorial velocity with a two-dimensional ESTER model. The values in boldface are the ones which were adjusted so that the other parameters (output of ESTER) would match those of \citetads{2007Sci...317..342M}}
\begin{tabular}{c | c | c} 
Parameters & \citetads{2007Sci...317..342M} & ESTER model\\[0.5ex]
\hline
$M$ (\msun)                & 1.791             & \textbf{1.65}\\
$\Tpol$ (K)                & 8450 $\pm$ 140    & 8450\\
$\Teq$ (K)                 & 6860 $\pm$ 150    & 6849\\
$\Rpol$ (\rsun)            & 1.634 $\pm$ 0.011 & 1.627\\
$\Req$ (\rsun)             & 2.029 $\pm$ 0.007 & 2.027\\
$\veq$ (\kms)              & 285.5 $\pm$ 6     & 274\\
$Z$                        &                   & \textbf{0.014}\\
$[\mathrm{M}/\mathrm{H}]$  & -0.2              & \\
$\Xc$                      &    --             & \textbf{0.35}
\end{tabular}
\label{table:first estimate}
\end{table}


\section{Atmospheric models for ESTER models} \label{section:Atmosphere}

The analysis of interferometric and spectroscopic data requires monochromatic intensity maps of the surface of the star from which the complex visibility will be computed. The interferometric observables and the spectrum will then be extracted from it. To obtain these intensity maps, model atmosphere codes coupled with radiative transfer codes must be used.

\subsection{Atmosphere models}

Among the different existing atmosphere codes, we decided to use models from the PHOENIX code\footnote{PHOENIX atmosphere models, spectra and specific intensities are available at \url{http://phoenix.astro.physik.uni-goettingen.de/}.} \citepads{2013A&A...553A...6H}, as it is the code that best matches our needs in terms of surface parameters (see Table \ref{table:first estimate}), spectral range, and specific intensity.

Indeed, these spherically symmetric atmospheres were used to produce specific intensities (low resolution, but for different values of the emission angle) and spectra (high resolution), from 500 to 26000 \angstrom, which suits our study since we are analysing observational data in the visible range (ELODIE), the H band (PIONIER), and the K band (GRAVITY). Models computed for 0.5 $< \logg <$ 6.0, 2300\kelvin < $\Teff$ < 12000\kelvin\ and -1.0 < [M/H] < 1.0 were selected from the PHOENIX online library.


To compute the intensity maps, as seen from Earth, we need the specific intensities as a function of the emission angle $I(\mu)$. These are available in the PHOENIX database, but only with a 1 \angstrom\ step. This kind of resolution is sufficient to analyse PIONIER and GRAVITY data, but as most lines are blended together by rapid rotation, initial high resolution spectra are needed to obtain an accurate broadened spectrum in the visible, which we need for the analysis of our spectroscopic data. Such high resolution spectra are only available in the database as fluxes, independent of the direction of emission. To compute the desired quantity, we used  \citetads{2000A&A...363.1081C}'s expression of limb darkening (Eq. 6), using the $a_k$-coefficients found in the dedicated Vizier catalogue (\citeads{2018A&A...618A..20C}, most recent catalogue for the wavelengths of interest). The expression 
\begin{equation}
    I_\lambda(\mu) = I_\lambda(\mu=1)\left[1-\sum\limits_{k=1}^4 a_k\left(1-\mu^{\frac{k}{2}}\right)\right],
    \label{eq:Claret}
\end{equation}
can be coupled with
\begin{equation}
    F_\lambda = 2\pi \int\limits_0^1 I_\lambda(\mu)\mu d\mu
\end{equation}
to yield
\begin{equation}
    F_\lambda = I_\lambda(\mu=1)\pi\left[1-\sum\limits_{k=1}^4 a_k\frac{k}{k+4}\right].
\end{equation}
Thus $I_\lambda(\mu=1)$ can be obtained from the flux $F_\lambda$, followed by $I(\mu)$, that is the specific intensity emitted in any direction. Let us point out that this method, based on band-integrated limb darkening coefficients, can only be approximate, especially if some thermal convection is present at the surface \citepads{1982ARA&A..20...61D}. A more precise method is however premature at this stage of the investigations.

\subsection{Linking interior and atmosphere models : ESTERIAS}

Once a complete grid of stellar models and a grid of atmosphere models are made, we use them both to create intensity maps. This is done with ESTERIAS (ESTER for Interferometry, Asteroseismology, Spectroscopy), the Python tool we developed to compute interferometric and spectroscopic observables with ESTER.

\subsubsection{The stellar surface grid}
A grid is used to represent the surface of the star, dividing it in surface elements $\mathrm{d}S$. We chose to construct the grid as `rings' of constant co-latitude $\theta$, each point of a ring being defined by a $\phi$ value. In order to do this, we impose the number of rings ($\theta$ values) from pole to pole, and the number of azimuthal points ($\phi$ values) of the first ring (from the pole). The radial distance (function of $\theta$) corresponding to every ring (the star being axisymmetric) must then be interpolated from the ESTER model over the $\theta$ values chosen for the grid. Indeed, ESTER models only require 20 grid points in latitude to achieve a precise solution. However, this is not enough for an accurate representation of the stellar spectra. Fortunately, the spectral representation with spherical harmonics allows the user to get the value of any parameter at any co-latitude, without losing numerical accuracy. Some tests showed us that the relative difference between a model computed with $N_\theta=100$ $\theta$-values and parameters interpolated on these 100 co-latitudes from a model made with only $N_\theta=10$ never exceeded 10$^{-4}$.

Once the `radius' and co-latitude of each ring are determined, the number of azimuthal
points must be set. To determine the appropriate numbers of points per ring, we
first imposed a number of points $N_{\phi_0}$ on the first ring. The area $dS_0$ of these $N_{\phi_0}$ 
surface elements is then computed and is identical for all points on the ring, as $dS$ only depends on 
$\theta$, $\mathrm{d}\theta$ and $\mathrm{d}\phi$, see Eq. \ref{eq:ds}. Finally, $N_\phi$ is computed for each ring such that $\mathrm{d}S$ is as close to $\mathrm{d}S_0$ as possible.
This is done using equation (A.51) of \citetads{2016JCoPh.318..277R} :
\begin{equation}
\mathrm{d}S = r^2\sqrt{1+\cfrac{{r_\theta}^2}{r^2}}\,\sin\theta \,\mathrm{d}\theta \mathrm\,{d}\phi,
\label{eq:ds}
\end{equation}
with $\mathrm{d}S$ the surface element area, and $r$ and $r_\theta$ the radial coordinate and its
derivative with respect to the $\theta$ coordinate.

Apart from $r$ and $r_\theta$, the relevant stellar surface parameters that must be extracted from the
models and interpolated on our $\theta$ values are: $\Teff$, the effective temperature; $\logg$,
the logarithm of the effective gravity; and $\Omega$, the angular velocity. The linear 
velocity $v_\phi = r(\theta) \Omega(\theta) \sin\theta$ is then computed.

We are thus left with a set of points, each associated with a set of ($r(\theta), \theta, \phi$) coordinates, and physical parameters ($\Teff(\theta), \logg(\theta), \Omega(\theta)$), that define our star surface.

\subsubsection{The visible grid}

\begin{figure}[t]
\centering
\includegraphics[width=0.45\textwidth]{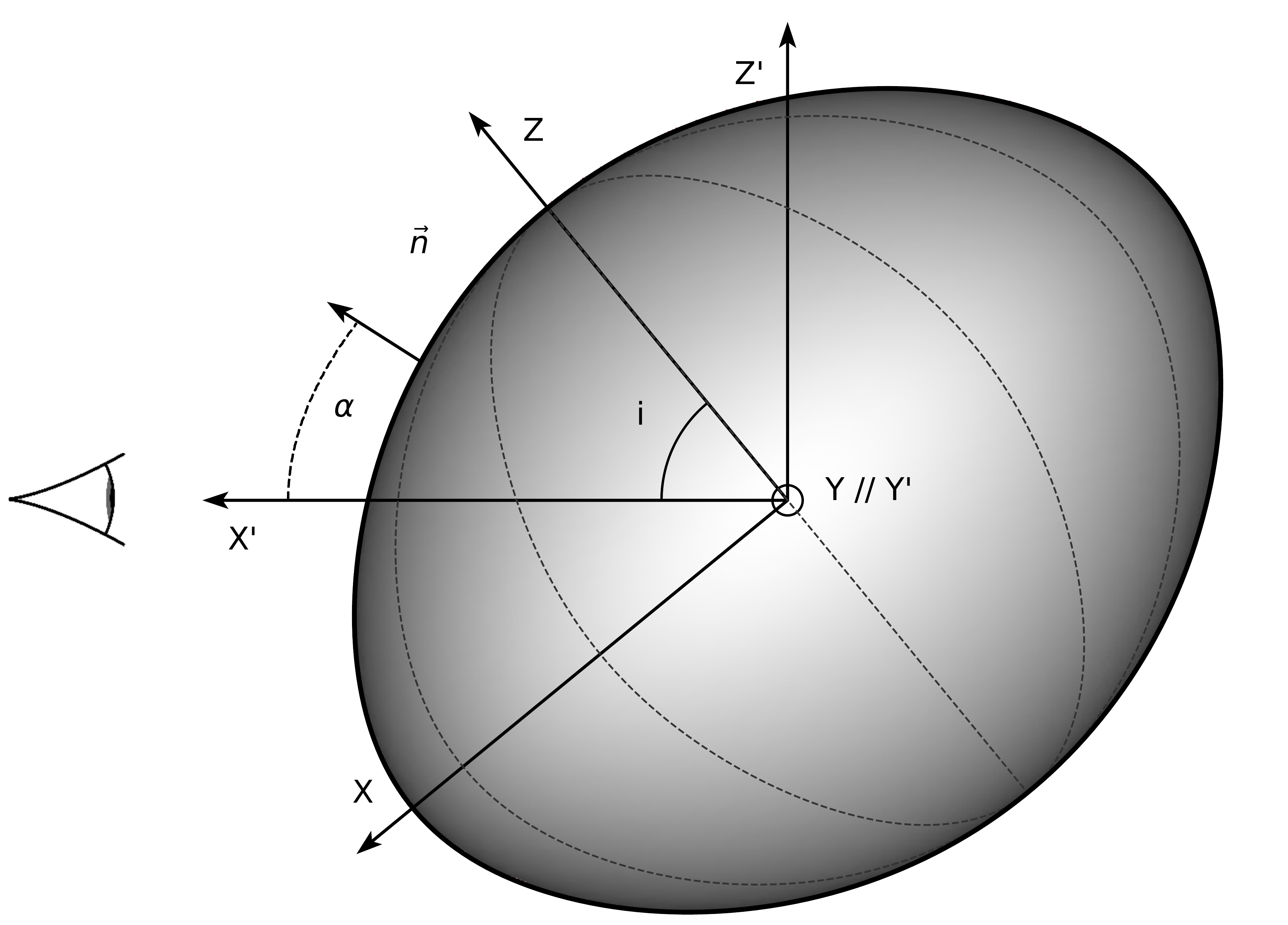}
\caption{Side view of a flattened star. The primed coordinates are the Cartesian coordinates used to describe the plane of the sky, as seen from Earth. (\textbf{Y', Z'}) are the axes of the sky plane, and $\vec{X'}$ the line of sight. The non-primed coordinates are linked to the star, and can be obtained by rotating the primed coordinates around the \textbf{Y'} axis so that the rotation axis $\vec{Z}$ is at an inclination angle $i$ from the line of sight. $\alpha$ is the angle between the normal $\vec{n}$ to the surface at an arbitrary point on the stellar surface and the line of sight. We recall that $\mu=\cos\alpha$ (cf. Eq.~\ref{eq:mu}). The star is shaded based on a 3D rendering of its shape, and not on physical effects such as gravity or limb-darkening. Other figures such as Figs. \ref{image:temp_map} and \ref{image:int_map} show such effects.}
\label{image:mu}
\end{figure}

Once the surface grid is determined, it can be used to generate a number of `visible grids', representing the visible side of the star, as viewed from an earthbound observer, for any value of the inclination $i$ of the star's rotation axis. For this we associate with each point a parameter $\mu$ which is the cosine of the angle $\alpha$ between the normal to the surface and the line of sight (see Fig. \ref{image:mu}):

\begin{align}
\mu &= \cos\alpha \nonumber\\
    &= \cfrac{\left(r\cos\theta +r_\theta \sin\theta \right)\cos i+\left(r\sin\theta -r_\theta \cos\theta\right)\cos\phi\sin i}{\left[r^2+r_{\theta}^2\right]^{1/2}}.
\label{eq:mu}
\end{align}
We then keep only the points for which $\mu \geq 0$.

Once the `visible grid' is ready, we can compute the geometrical parameters $\vec{Y'}$, $\vec{Z'}$ and $\vec{\mathrm{d}s_{\mathrm{proj}}}$, respectively the two Cartesian coordinates projected onto the plane of the sky (see Fig. \ref{image:mu}) and the projected surface element, as well as the linear velocity projected onto the line of sight $v_{\mathrm{proj}}$. We find that: 

\begin{align}
Y'\phantom{_{\mathrm{proj}}} ={}& r\ \sin\theta\ \sin\phi,\\
Z'\phantom{_{\mathrm{proj}}} ={}& r\ (\cos\theta\ \sin i - \sin\theta\ \cos\phi\ \cos i), \nonumber \\
\begin{split}
	\mathrm{d}s_{\mathrm{proj}}  ={}& r \sin\theta \mathrm{d}\theta \mathrm{d}\phi\ (\cos\phi \sin i\left(r\sin\theta -r_\theta \cos\theta\right)\\
	&\qquad\qquad\quad+\cos i\left(r\ \cos\theta +r_\theta \sin\theta \right)),
\end{split}\nonumber\\
v_{\mathrm{proj}}\phantom{d}  ={}& r\ \Omega \sin\theta \sin\phi \sin i.\nonumber
\end{align}

At this step in the process, we have a two-dimensional representation of the stellar surface (a 2D projection of a 3D object actually), giving a map of its shape, temperature, gravity, and rotation velocity.

\subsubsection{The intensity map}

Several operations (mainly interpolations) are necessary to get the final intensity maps from the specific intensity files. First it is necessary to associate the original intensity $I_\lambda$ from the files, which were computed for fixed values of effective temperature and gravity, with every point on the grid. 
Then, for each $\mu$ associated with each visible grid surface element, the $I_\lambda$ are interpolated from the original files, or computed with Claret's formula (Eq. \ref{eq:Claret}).
Finally, the Doppler shift due to the rotation of the star is computed. We thus get one intensity map per wavelength value (monochromatic intensity maps), for which the several observables (e.g. interferometric, spectroscopic) can be computed.

\subsubsection{Observable quantities}\label{observables}

For the interferometric data, the two components of the sky-projected baselines ($B_x, B_y$), the squared visibilities and closure phases with their associated errors, and the wavelengths are extracted, and the spatial frequencies $\left(u = \frac{B_x}{\lambda},
v = \frac{B_y}{\lambda}\right)$ are computed. Here, the $x$ and $y$ subscripts refer to the cartesian sky (angular) coordinates, with $x$ positive towards the east and $y$ to the north. A transformation taking into account the position angle of the star (PA, the angle between the northern direction and the axis of rotation projected onto the plane of the sky, counted positive to the east) must be applied on Y' and Z' (from Fig. \ref{image:mu}) to get $x$ and $y$ (e.g. in Eq. \ref{eq:I_u_v}). As the star is made of a series of rings of different
sizes, with non-uniform spatial discretisation, a Fast Fourier Transform algorithm cannot
be used. We thus computed the discrete Fourier Transform via the classical formulas :
\begin{align}
\tilde{I_\lambda}(u, v) &= \iint I_\lambda(x, y) e^{-2i\pi (xu + yv)} \mathrm{d}x \mathrm{d}y\\
&\simeq \sum\limits_{j} I_\lambda(x_j, y_j) e^{-2i\pi (x_ju + y_jv)} \frac{\mathrm{d}S_{\mathrm{proj}}}{d^2},
\label{eq:I_u_v}
\end{align}
$d$ being the distance to the star. The complex visibility, at each $\lambda$, is then given by
\begin{equation}
V = \frac{\tilde{I}(u, v)}{\tilde{I}(0, 0)},
\end{equation}
with $\tilde{I}(0, 0)$ the flux integrated over the visible surface of the star (observed flux).
The squared visibilities and closure phases are computed from the complex visibilities with
\begin{equation*}
V^2 =\ \mid\!V\!\mid^2 , \qquad \Psi_{\textrm{cl}} = \Psi_{12} + \Psi_{23} - \Psi_{13}.
\end{equation*}
$\Psi\sub{ij}$ being the phase of the complex visibility computed at the projected baseline defined by telescopes i and j, for i $\neq$ j, and i, j = 1, 2, 3 (three telescopes).

\section{Model Fitting} \label{section:Model Fitting}

We apply a model-fitting approach to find the parameters that best represent the star, using the emcee code \citepads{2013PASP..125..306F}. This Python implementation of the Markov Chain Monte Carlo (MCMC) method allows the user to easily tweak the parameters of the fit: the number of values to test at every step, the number of steps, the priors to apply to constrain our fit, etc. The MCMC method consists in drawing a constant number of random `walkers', i.e sets of values for the free parameters of the problem, within the allowed range for each parameter, and computing the posterior probability density function associated with each set. A new ensemble of values is then randomly drawn, and each walker will be replaced by the new value with some probability, the probability being higher if the new value gives a better fit than the previous one. The process is then repeated for a preset number of steps, or until convergence is met. Indeed, as the process proceeds, the walkers will be more and more closely gathered around the best-fitting solution(s). The randomness of the process enables to detect and avoid local minima, at the condition that the parameter space be sufficiently sampled (by choosing a high enough number of walkers, and the right parameter space domain).

Unfortunately, as some parameters become more `extreme' (mass below 2 \msun, low or high metallicity, high rotation velocity, etc.), ESTER will fail to create the model if the input model is not really close to the output solution.
A step-by-step approach, creating intermediate models before reaching the desired one, can be used, but not as part of an MCMC model-fitting. Indeed, as there is no way to know beforehand whether ESTER will converge on a solution, this step-by-step approach has to be done manually, changing the increment, or the input parameter to increment, if needed.
Furthermore, computing ESTER models is very time-consuming, as one model takes between 30 seconds and several minutes to produce. This may be fast for the computation of the full two-dimensional structure of a star, but isn't appropriate for a model-fitting needing the computation of several tens of thousands of models.

We thus decided to first create a grid of stellar models with ESTER, with fixed increments in parameter values. The parameter space covered was chosen by taking into account either the parameter values found in the literature (when available) or various assumptions concerning plausible ranges for other parameters ($\Xc$ for example). Then, ESTERIAS  was made so that when random values of parameters are drawn by the MCMC routine, the parameters which are relevant for our star surface (effective temperature and gravity, radius, and rotation velocity) are linearly interpolated for these values from the closest models in our grid (selected via Delaunay triangulation).

We present in the next section the different attempts at determining the following parameters of Altair : mass ($M$), equatorial radius ($\Req$), angular velocity ($\Omega$, expressed as a fraction of the Keplerian angular velocity\footnote{The true critical velocity $\Omega\sub{c}$ is indeed unknown unless the actual (critical) model is computed. Often the critical angular velocity of the Roche model is used as a reference, but it is a mere transformation of the actual Keplerian velocity, which we use.  Further details may be found in the appendix of \citetads{2016LNP...914..101R}.}, $\OmegaK = \sqrt{GM/\Req^3}$), metallicity ($Z$), hydrogen mass fraction in the core ($\Xc$), inclination ($i$), position angle ($PA$), and the metallicity of the atmosphere models ([M/H]). The envelope hydrogen content $X$ is also considered, as it turns out it plays an important role in the convergence towards a physically realistic solution.

\section{Results} \label{section:Results}

\subsection{Fitting interferometric and spectroscopic data}

First, our attempts to determine Altair's parameters by fitting interferometric and spectroscopic data are described. Additional constraints obtained through other methods are presented in the following sections.

\subsubsection{\texorpdfstring{$\omega$-model}{omega-model}}

Initially, an attempt was made to constrain simultaneously all the parameters previously mentioned, using ESTER models and only interferometric data. The $\chi^2$ was computed in a way that the two sets of data (PIONIER and GRAVITY) had the same weight. Convergence could not be achieved, as there were several correlations between parameters, such as $M$, $Z$ and $\Xc$, or $\Omega$ and $i$, and constraining all seven parameters with only interferometry was impossible. To disentangle this, we decided to study the parameters in smaller groups. As ESTER models need all the parameters as input, and are quite heavy to use, they were replaced by the so-called `$\omega$-model' \citepads{2011A&A...533A..43E,2016LNP...914..101R} for this first step. This gravity-darkening model provides all of the relevant surface parameters ($\Teff$, $\geff$, $r$, etc.) without going through the difficult process of computing the whole internal structure of the star (a Roche-model is used to compute the effective gravity). It stems from simple assumptions, namely energy is conserved, there are no energy sources in the envelope, and the radiative flux is anti-parallel to the effective gravity. Thus, the resulting models, while being in almost-perfect agreement with ESTER models (see Fig. 5 from \citetads{2016LNP...914..101R} and Fig. 3 from \citetads{2011A&A...533A..43E}), do not allow us to adjust internal parameters such as the metallicity or hydrogen fraction in the core. They are still suitable for studying the shape of the star and the distribution of flux on the surface though, making it perfect for the analysis of the interferometric data, which is mostly sensitive to these two signatures of rotation. Such an analysis was already done by \citetads{2018A&A...619A.167D}, who succeeded in reproducing the interferometric observables of the star \object{Sargas}, an evolved 5.1~\msun\ rapidly rotating star, with the use of the $\omega$-model. 

\begin{figure*}[t]
\centering
  \includegraphics[width=0.8\textwidth]{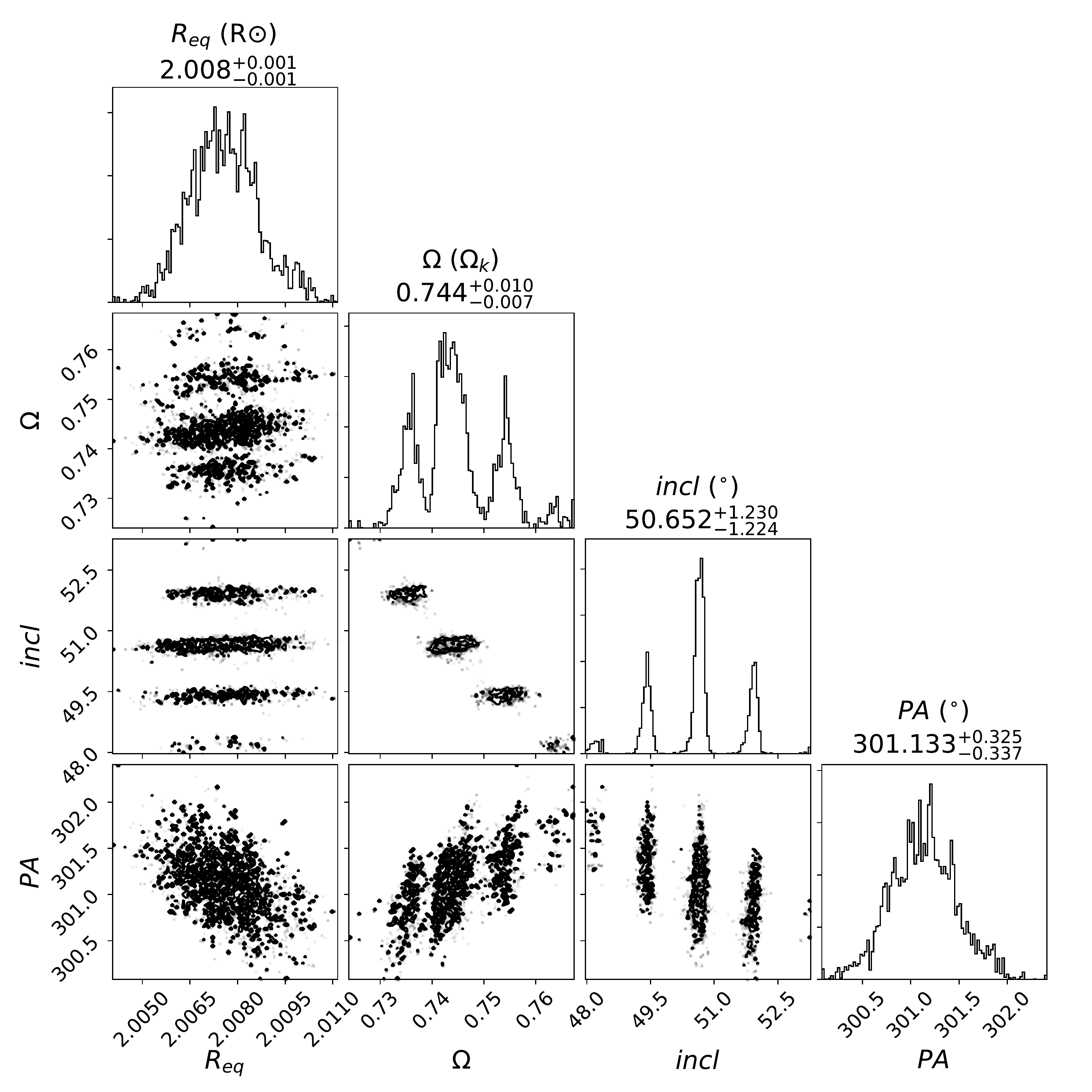}
	\caption{Corner plot showing the convergence of the MCMC model-fitting for $R_{\textrm{eq}}$, $\Omega$, $i$, and $PA$, using interferometric observations (PIONIER + GRAVITY) and the $\omega$-model. The run was made with 200 steps and 100 walkers. Only the last 50\% steps are displayed here as convergence was met within the first 100 steps ("burn-in" phase)}. These results are listed in Table \ref{table:final_model}.
\label{image:corner_first_four}
\end{figure*}

The parameters we are constraining with interferometric data and the $\omega$-model are the equatorial radius, $\Req$, the angular velocity, $\Omega$ (again, here, as a fraction of the Keplerian equatorial velocity), the inclination, $i$, and the position angle, $PA$. The mass, $M$, and luminosity, $L$, must also be given as input, but as they only affect the scale of $\Teff$ and $\geff$, not their distribution, the interferometric observables won't be sensitive to those. We used $M$ = 1.8 \msun\ and $L$ = 10.6 \lsun, both taken from \citetads{2006ApJ...636.1087P}. Altair's distance was fixed to 5.13 pc, from the HIPPARCOS parallax of 194.95 $\pm$ 0.57 milliarcsecond (mas).
Clear convergence was obtained for $\Req$ and $PA$ (see Fig. \ref{image:corner_first_four}). For $\Omega$ and $i$, convergence is still achieved, but an effect due to the resolution of the intensity maps is clearly visible. This is discussed in Sect. \ref{section:Discussion}. The envelope of the peaks in the histograms of Fig. \ref{image:corner_first_four} is of Gaussian shape, with a relatively small width (judging from the $\pm 1\sigma$ values).

\begin{figure}[t]
\centering
  \includegraphics[width=0.48\textwidth]{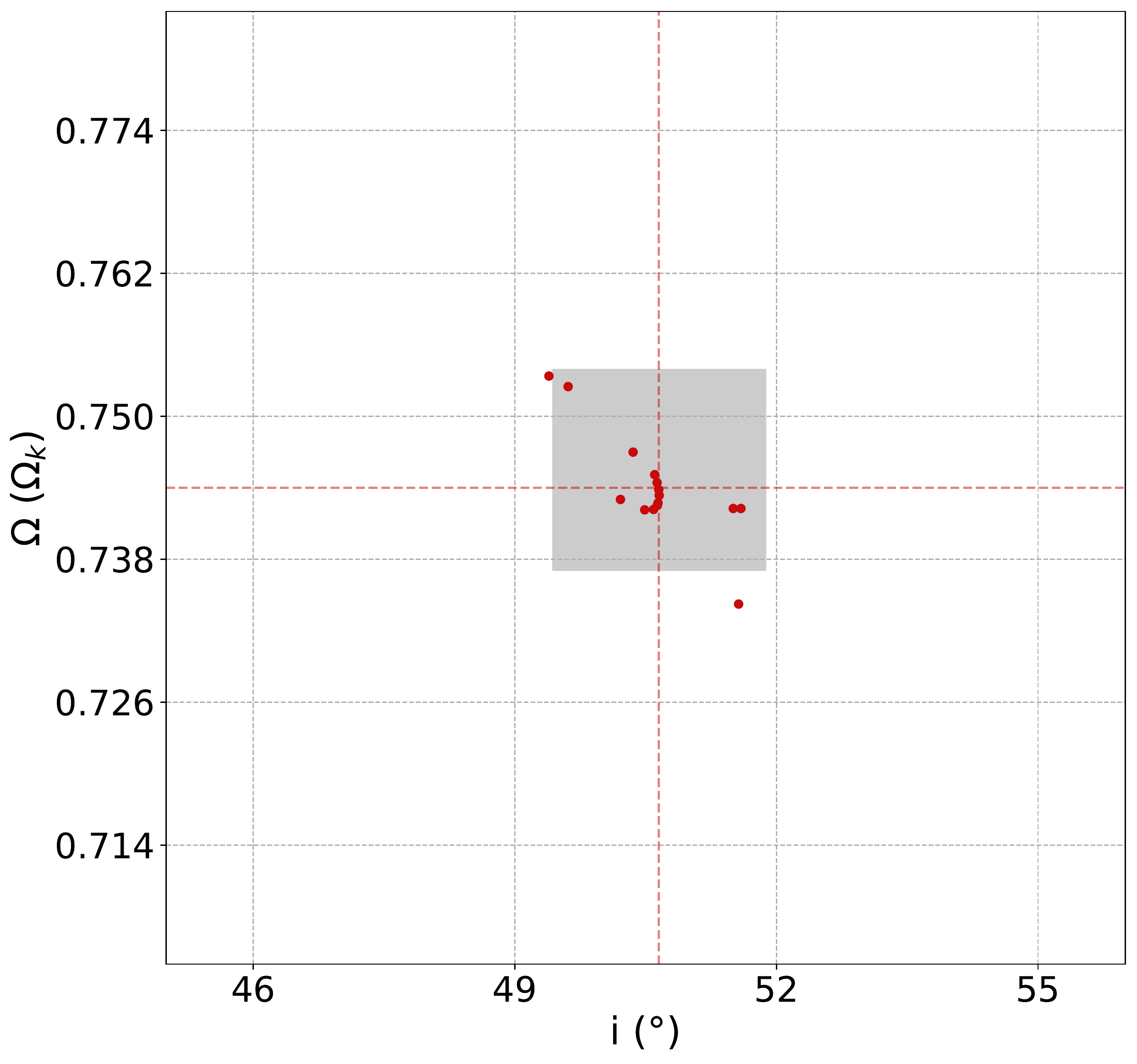}
	\caption{Positions in the ($i$, $\Omega$) plane of the $\chi^2$ minima when fitting interferometric data with $\Omega$ and $i$ as free parameters, using $M = 1.80$\msun, and several values of [M/H] for the atmosphere models. [M/H] goes from -0.5 to 0.9 with a 0.1 step. The grey rectangle shows the area enclosed within the error on $i$ and $\Omega$ shown in Fig. \ref{image:corner_first_four}.}
\label{image:M_H_coords}
\end{figure}

In the above MCMC model-fitting, we used model atmospheres with a fixed value of the metallicity, namely [M/H]=0.3. However, one may wonder what impact this metallicity can have on the fitting process.  Accordingly, we computed the $\chi^2$ value of the fit to the interferometric data with only $\Omega$ and $i$ as free parameters, for different values of the metallicity of the atmosphere models. We recall that the metallicity of the PHOENIX atmospheres is defined as
\begin{equation}
    \mathrm{[M/H]} = \mathrm{\log\left(\frac{n_M}{n_H}\right) - \log\left(\frac{n_M}{n_H}\right)_\sun},
\end{equation}
where $n$ is the number density of elements in the star, and $M$ the sum of all metals. As seen in Fig. \ref{image:M_H_coords}, showing the $i$ and $\Omega$ coordinates of the $\chi^2$ minima, the location of the best fit (in terms of ($\Omega$, $i$) coordinates) hardly changes. This means that the fitting of the interferometric data is nearly independent of this parameter. Spectroscopy is the go-to method to determine the metallicity of a star's atmosphere. Yet, only an extensive analysis of Altair's spectrum would allow us to get an accurate estimate of its composition. Indeed, at such a high rotation rate, neighbouring metallic lines are blended together. Thus, the integrated spectrum will not only depend on the lines' depth, but also on the list of lines included in the computation of the opacities and on the relative abundances of all elements in the atmosphere. These effects may be responsible for a mismatch between theoretical and observed spectra. Such work needs a dedicated study and is beyond the scope of this paper. We settled on using the line of Mg\textsc{II} at 4481\angstrom\  to get an appropriate atmosphere metallicity for the next steps. This line is strong and isolated enough that blending is not too much of an issue, and was deemed by \citetads{2009LNP...765..207R} as a valid candidate for $\vsini$ measurement in an A7-type star, which will give us one more constraint on this parameter. The line depth will give us an estimate on [M/H], keeping in mind that the abundance ratios in PHOENIX atmospheres are solar-like.
We computed a high resolution spectrum in the range of the Mg\textsc{II} line, and found that a value of [M/H] = +0.45 gave the best concordance with the observed spectrum (the integrated theoretical line is identical to the one obtained with our best ESTER model, shown in Fig. \ref{image:spectrum}). We also computed $\vsini$, and found $\sim$238 \kms, at the limits of the range 227 $\pm$ 11 \kms\ found by \citetads{2004A&A...428..199R}.

\subsubsection{ESTER models} \label{section:m_z_xc}
We now want to determine the mass $M$, the metallicity $Z$, and hydrogen mass fractions in the envelope and core, $X$ and $\Xc$ respectively. ESTER models are used, with $i$, $PA$ and $\Omega$ fixed to the previously obtained values. Interferometric and spectroscopic data are analysed simultaneously.

\begin{figure*}[t]
\centering
  \includegraphics[width=\textwidth]{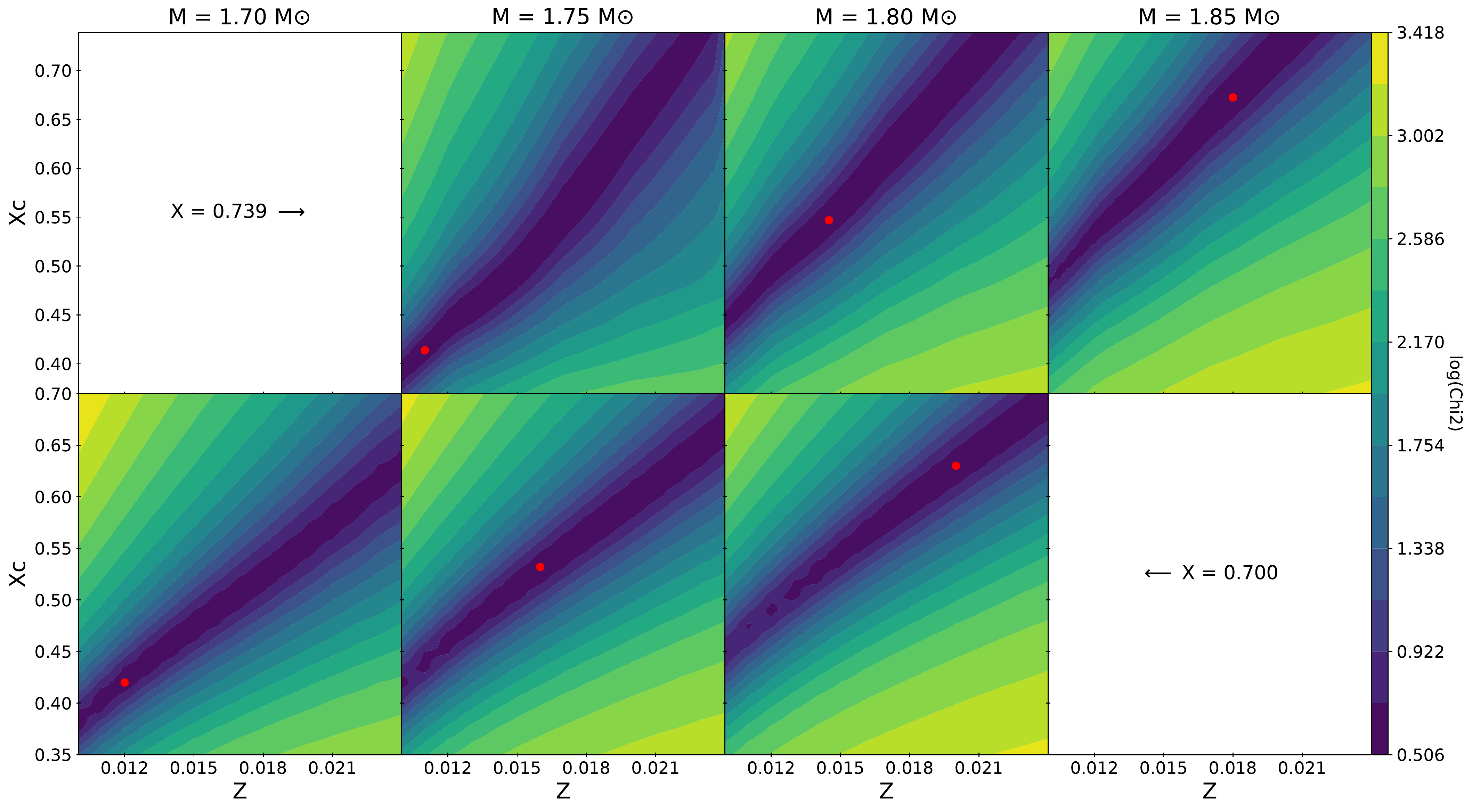}
	\caption{$\chi^2$ maps of the fitting of ESTER models to both interferometric and spectroscopic data. The upper and lower rows correspond to $X=0.739$ and $0.700$, respectively.  For each mass and $X$ value, only $Z$ and $\Xc$ vary, while every other parameter is fixed to the values previously obtained (see Fig. \ref{image:corner_first_four}). The decimal logarithm of the reduced $\chi^2$ is shown in colour. The red dot in each subpanel corresponds to the minimum $\chi^2$ value.}
\label{image:z_xc_intersection}
\end{figure*}

\begin{figure*}[t]
\centering
  \includegraphics[width=0.9\textwidth]{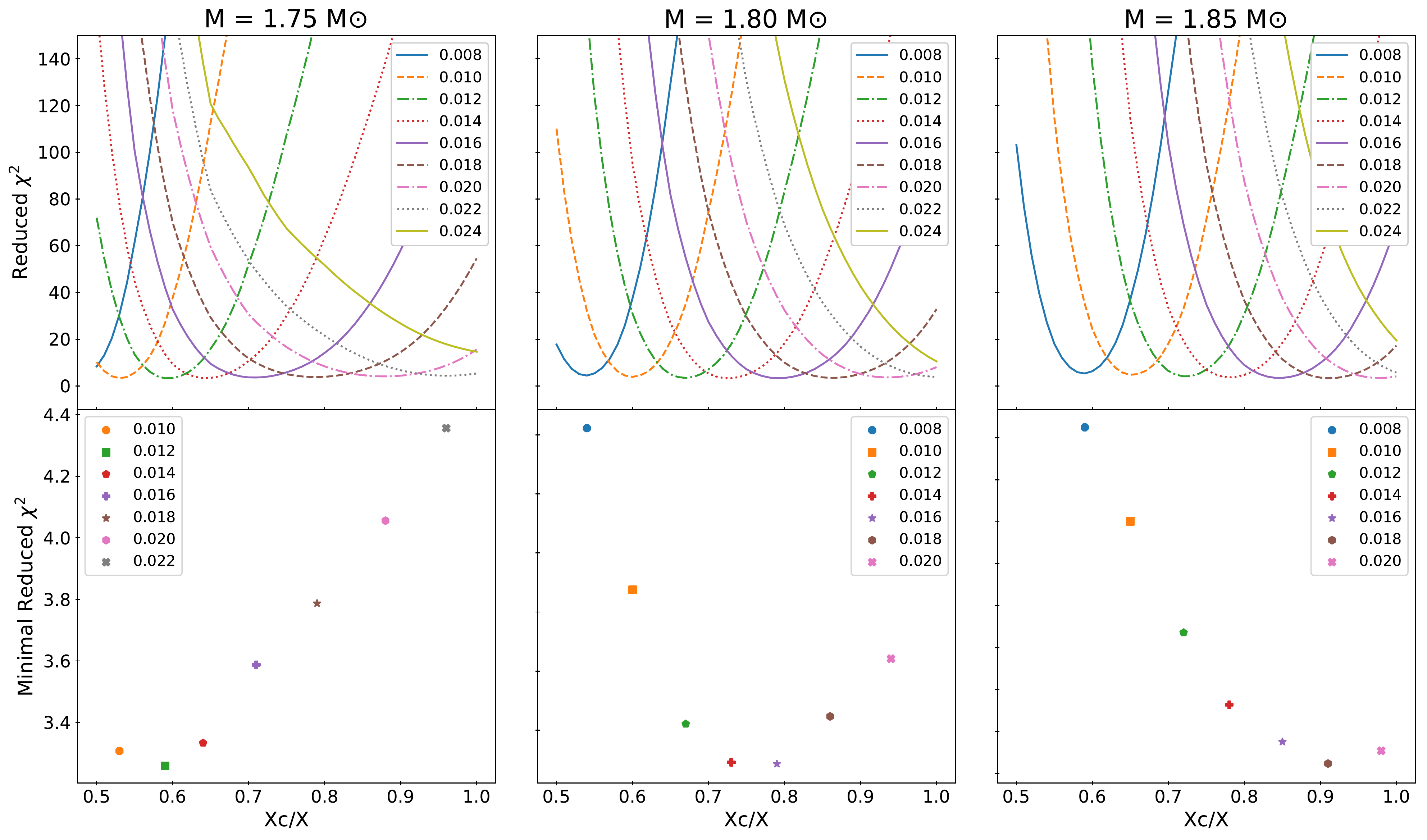}
	\caption{$\chi^2$ cuts at fixed $Z$ values (upper row), and along the valley with low $\chi^2$ values (lower row), for $X = 0.739$. The points in the lower row correspond to the minima in the upper row. As can be seen, a definite minimum appears even within the valley with low $\chi^2$ values. $Z$ values are shown in legend, and only the values for which the optimal $\Xc/X$ is in the range [0.5-1.0] are shown in the lower row.}
\label{image:z_xc_chi2_curves}
\end{figure*}

The metallicity of the stellar model must be addressed here.  Ideally, this metallicity should match that of the model atmospheres unless some physical phenomena, such as diffusion, were to lead to a different composition in the surface layers.  In the present case, we treat the two metallicities as independent parameters as they are subject to different constraints.  Indeed, the atmospheric metallicity is especially sensitive to spectroscopic constraints as shown above.  In contrast, the stellar model bulk metallicity is subject to both interferometric and spectroscopic constraints due to its impact on the stellar structure. Furthermore, it is correlated with $M$, $X$, and $\Xc$. Indeed, when the mass decreases, the size of the star also decreases; but increasing $Z$ increases its size, thereby compensating the effects of the mass decrease. Decreasing $\Xc$ (for a fixed $X$) mimics a more evolved star and, therefore, also increases its size. The dependence of stellar size on $X$ is more complex. In order to circumvent these difficulties, we decided to separate the parameters, and search for $Z$ and $\Xc$ simultaneously, for different values of the mass and envelope hydrogen mass fraction. We chose two values of $X$, namely $X=0.700$ \citepads{1993oee..conf...15G} and $X=0.739$ \citepads{2005ASPC..336...25A}. 
Fig. \ref{image:z_xc_intersection} shows the resulting $\chi^2$ maps. For both values of $X$, $\Xc$ is such that $0.5 < \Xc/X < 1.0$. We started with $X=0.700$, covering masses from 1.70 to 1.80 \msun, expecting Altair to be in this range, between our first estimate of Sect. \ref{section:Altair} and \citetads{2006ApJ...636.1087P}'s value. Then, as an asteroseismic study of these models favoured higher masses (see Sect. \ref{section:astero}), we opted for masses between 1.75 and 1.85 \msun\ for $X=0.739$. A mass of 1.90 \msun\ or higher would lead to a higher hydrogen mass fraction in the core than in the envelope, which wouldn't make sense with regard to current theories of stellar formation and evolution. Fig. \ref{image:z_xc_chi2_curves} shows cuts at fixed $Z$ values (upper row), and along the valley with low $\chi^2$ values (lower row), for $X = 0.739$. The figure would be the same for $X=0.700$, with $\chi^2$ minima shifted towards higher values of $\Xc/X$.

For a fixed value of $X$, the loci (in ($Z$, $\Xc$) coordinates) of the $\chi^2$ minima as a function of the mass (red points in Fig. \ref{image:z_xc_intersection}) show almost linear relations for $Z = f(M)$ and $\Xc = f(M)$, but a degree 2 polynomial seems to fit them better. We thus obtain two degree 2 polynomials as a function of $M$ for $Z$, and two for $\Xc$. We then do a linear interpolation on the coefficients of these polynomials between $X=0.700$ and $X=0.739$, and obtain a single expression for $Z$, as a function of both $M$ and $X$, and likewise for $\Xc$:
\begin{eqnarray}
    Z &\simeq& (0.0626\,X - 0.0494)\,M^2 \nonumber \\ 
      &  & + (-0.5922\,X + 0.5145)\,M  \nonumber  \\
      &  & + (0.7403\,X - 0.6602),
    \label{eq:correlation_z}
\end{eqnarray}
 and
 \begin{eqnarray}
     \Xc &\simeq& (-17.062\,X + 9.754)\,M^2 \nonumber  \\
         & & + (79.124\,X - 45.647)\,M \nonumber \\ 
         & & + (-88.777\,X + 52.334).
    \label{eq:correlation_xc}
 \end{eqnarray}
We may now check whether these relations allow us to retrieve the parameters we had found in Sect. \ref{section:Altair} by matching the temperature and radius of \citetads{2007Sci...317..342M}.
Extrapolating these relations to a lower mass of $M = 1.65$\msun\ and an $X$ value of 0.700, we get $Z \sim 0.008$, and $\Xc \sim 0.30$. This is somewhat different from our first estimate (see Fig. \ref{table:first estimate}). This is not surprising, as our results using both the $\omega$-model and ESTER give a slightly smaller size and broader surface temperature range than \citetads{2007Sci...317..342M} (this is discussed in Sect. \ref{section:Discussion}). 

If we can now obtain a value for the mass and hydrogen mass fraction through other means, we will get $Z$ and $\Xc$ at the same time. In what follows, we investigate what constraints can be placed on the mass.

\subsection{Spectral energy distribution} \label{section:SED}

\begin{figure}
\centering
  \includegraphics[width=0.48 \textwidth]{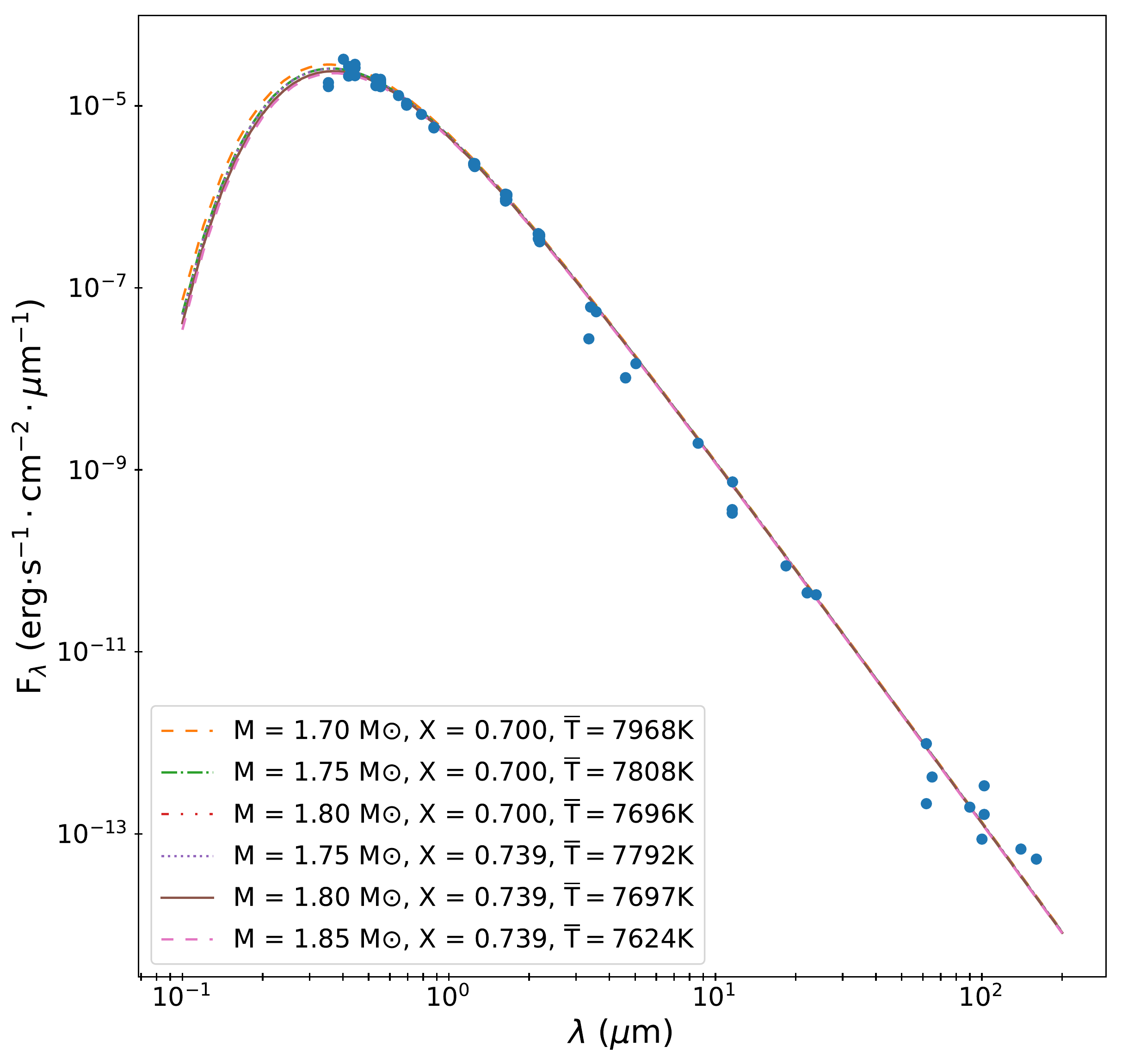}
	\caption{Spectral energy distribution produced in the direction of the observer (the inclination of the star being the one shown in Table \ref{table:final_model}), for different ESTER models with $X=0.700$ and $0.739$. $\overline{T}$ is the mean surface temperature of the models. The blue dots correspond to the observed SED data from the Vizier website.}
\label{image:SED}
\end{figure}

One solution to determine the mass is to compare the Spectral Energy Distribution (SED) of the model with the observed SED of Altair. The SED, which is a measure of the energy received on Earth from the star at different wavelengths, increases as the mass increases (as it mainly depends on the effective temperature of the star, and its radius). But this is only true when the other physical parameters of the star are kept constant. The correlation between $M$, $Z$, $X$, and $\Xc$ makes it so that the models which follow the previously found relation have a similar temperature range and distribution (this includes the size and shape of the model star), despite having a different mass. Their SED should thus be fairly similar, the higher-mass models nonetheless having a slightly lower mean temperature ($\sim$ 270\kelvin\ decrease from $M = 1.70$\msun\ to $M = 1.80$ \msun, for $X=0.700$, and $\sim$ 170\kelvin\  decrease from $M = 1.75$\msun\ to $M = 1.85$ \msun, for $X=0.739$).
To test this hypothesis, we computed the SED of three models for $X=0.700$ (corresponding to four different masses), and 3 models for $X=0.739$, with $Z$ and $\Xc$ following the relations previously found. The result is shown in Fig. \ref{image:SED}. The observational data used is available on Altair's page of the Vizier photometry viewer\footnote{Data points and their associated references can be found on: \url{http://vizier.u-strasbg.fr/vizier/sed/}}. Apart from the blue part of the curve, where the effect of the temperature difference is visible, the SED are nearly identical.

Comparing the projected rotation velocities of the models did not help, as the $\vsini$ found for these 4 ESTER models ranges from 229 to 238 \kms\ with increasing mass, still within the error bars of \citetads{2004A&A...428..199R}.

\subsection{The word of asteroseismology} \label{section:astero}

Since Altair is a $\delta$ Scuti, an alternate approach to constraining its mass is to model its observed pulsation spectrum.  Indeed, acoustic pulsation modes (p modes) potentially provide a tight constraint on the mean density \citepads[e.g.][]{2008A&A...481..449R, 2015ApJ...811L..29G}, which when combined with the volume, provides the mass.  $\delta$ Scuti type pulsations were discovered in Altair thanks to the star tracker on the WIRE satellite \citepads{2005ApJ...619.1072B}.  The pulsation frequencies and amplitudes are reproduced in Table~\ref{table:pulsations} for convenience, with the first seven being arranged in order of increasing frequency, while keeping the original indices of \citeads{2005ApJ...619.1072B}, who ordered the modes by decreasing amplitude. \citetads{2005A&A...438..633S} subsequently attempted to interpret Altair's pulsations using a second order perturbative approach to model the effects of rotation on the pulsations.  In the present study, we use a full 2D approach, both for stellar structure and pulsations, which is necessary for a star rotating as rapidly as Altair \citepads[e.g.][]{2006A&A...455..621R}.

One of the difficulties with which \citetads{2005A&A...438..633S} were confronted and which still remains in the current study is the fact that the mode identification, i.e. the correspondence between observed and theoretically calculated modes (as characterised through a set of quantum numbers), is unknown.  Rather than carrying out an exhaustive search for best fitting pulsation spectra using a $\chi^2$ minimisation, we prefer to make a few simplifying assumptions and see to what conclusions they lead.  First of all, the frequencies $f_1$, $f_2$, $f_7$, $f_3$, and $f_6$ form a regular pattern with subsequent frequencies being spaced by roughly $5$ or $2.5$ cycles/day.  Furthermore, their amplitudes alternate between higher and lower values.  A compelling interpretation is to assume that these are a sequence of island acoustic modes with the same $\tilde{\ell}$ and $m$ values and consecutive $\tilde{n}$ values, with a mode missing between $f_1$ and $f_2$ (see e.g. \citeads{2009A&A...500.1173L, 2012A&A...546A..11P} for an explanation on island modes and associated quantum numbers).  The alternating amplitudes could partially be explained by the fact that modes with even $\tilde{n}$ values are symmetric with respect to the equator whereas those with odd values are antisymmetric.  Indeed, for a given mode normalisation, this would lead to different apparent pulsation amplitudes (or mode visibilities) when integrating the intensity fluctuations over the visible disk because of differing degrees of cancellation.

\begin{table}[ht]
\centering
\caption{Altair's pulsation spectrum (reproduced from \citeads{2005ApJ...619.1072B}).}
\begin{tabular}{c c c} 
\hline
Mode & $\nu$ (c/d) & $A$ (ppm) \\
\hline
 $f_1$ & 15.768 & 420 \\
 $f_4$ & 15.989 & 195 \\
 $f_5$ & 16.183 & 140 \\
 $f_2$ & 20.785 & 377 \\
 $f_7$ & 23.280 & 108 \\
 $f_3$ & 25.952 & 245 \\
 $f_6$ & 28.408 & 123 \\
 $f_8$ &  2.570 & 104 \\
 $f_9$ &  3.526 &  92 \\
\hline
\end{tabular}
\label{table:pulsations}
\end{table}

\begin{figure*}[ht]
\centering
\includegraphics[width=0.48\textwidth]{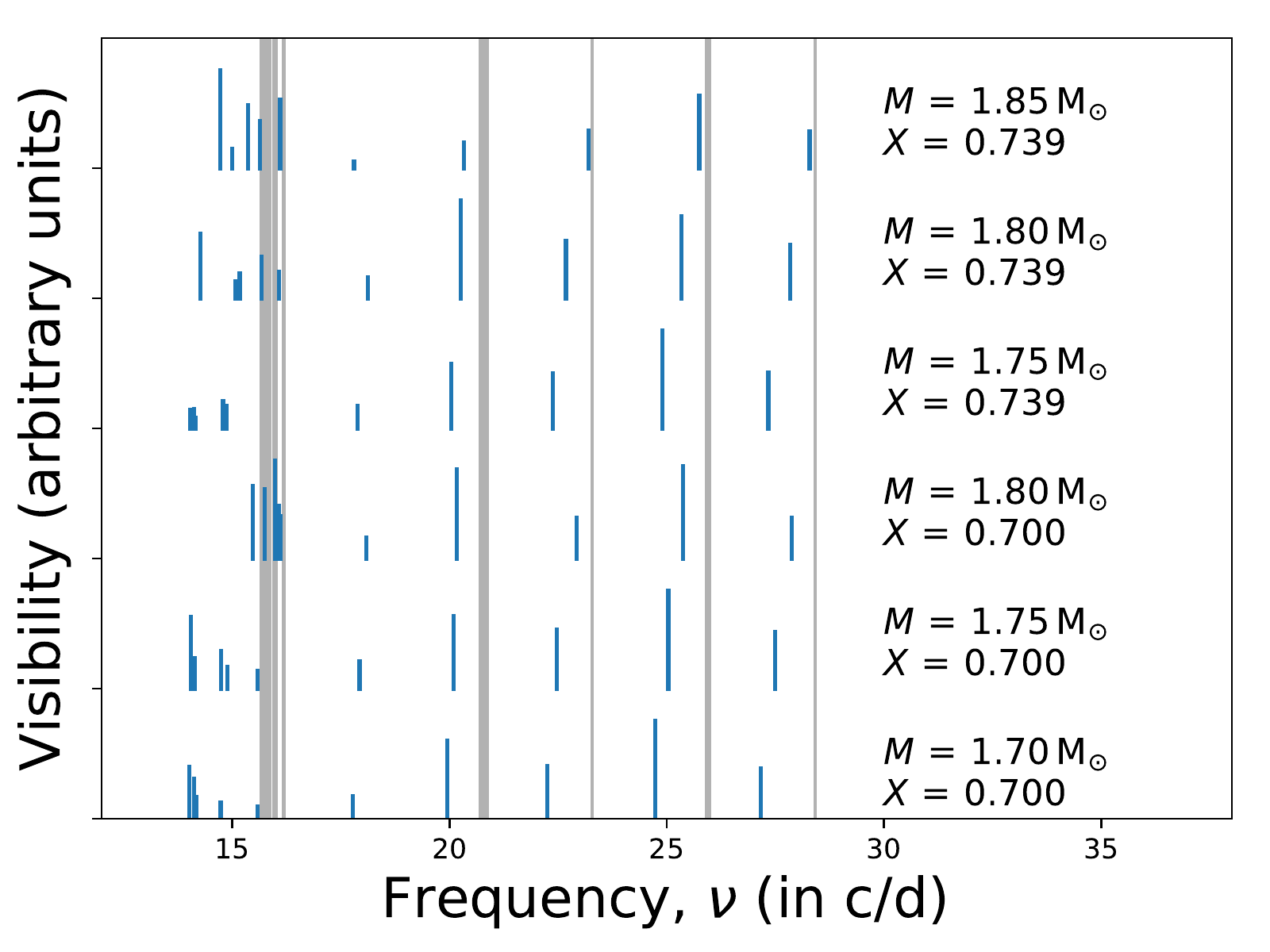}
\includegraphics[width=0.48\textwidth]{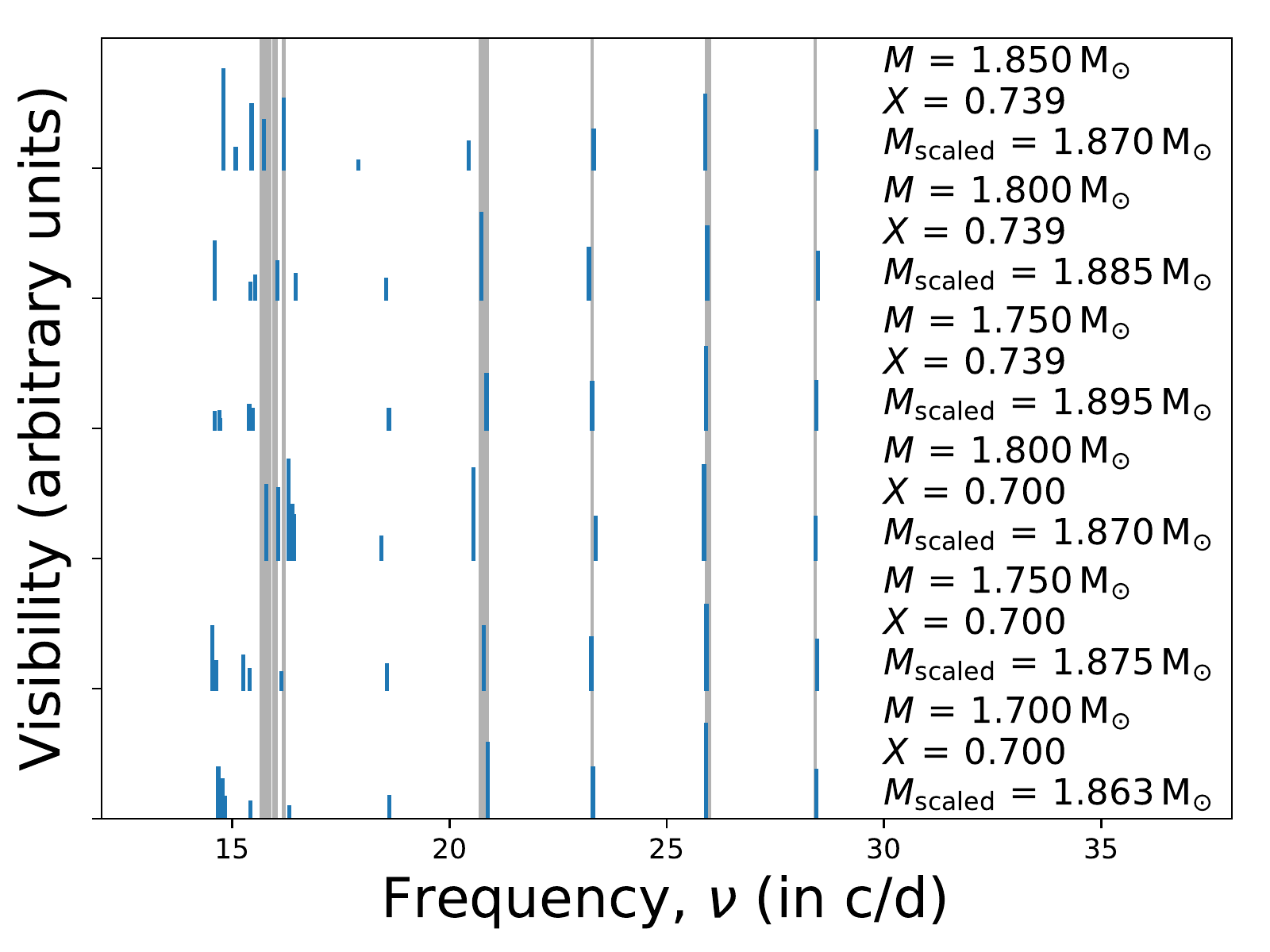}
\caption{Observed and theoretical pulsations for Altair.  The vertical grey lines that span the plots are the observed pulsations (modes $f_1$ to $f_7$).  Their thicknesses are proportional to the observed amplitudes.  The blue vertical segments are theoretical pulsations for six different ESTER models which satisfy Eqs. \ref{eq:correlation_z} and \ref{eq:correlation_xc}. The theoretical modes are successive $m=0$, $\tilde{\ell}=0$ (or equivalently $\ell=0$ and $1$) modes apart from the first mode (see text for details).  The lengths of these segments are proportional to the disk-integrated mode visibilities.  The left plot shows the theoretical frequencies whereas the right plot shows the same set of frequencies after applying a suitable homology scaling \citepads[e.g.][]{1990sse..book.....K}, thus leading to the scaled masses indicated on the figure (assuming the equatorial radius is fixed).}
\label{FigAstero}
\end{figure*}

Figure~\ref{FigAstero} shows a comparison between observed and theoretical frequencies for six models that satisfy the interferometric and spectroscopic constraints. For the theoretical modes, we used $(\tilde{\ell},\,m)=(0,0)$ which was the simplest assumption to make and leads to selecting the most visible modes.  In terms of spherical quantum numbers, these would be the rotating equivalents to $\ell=0$ and $\ell=1$ modes.  Furthermore, we calculated disk-integrated mode visibilities using the approach given in \citetads{2013A&A...550A..77R}, the inclination obtained from interferometry (i.e. $i=50.65^{\circ}$), and a photometric band deduced from Fig.~8 of \citetads{2004ESASP.538..205B}.  The modes were normalised such that the maximal Lagrangian displacement times the square of the frequency is kept constant.  In \citeads{2013A&A...550A..77R}, this was found to keep approximately constant visibilities for island modes with the same $(\ell,m)$ values and to penalise gravity modes which tend to have a low surface amplitude.  As can be seen, the alternating amplitudes are correctly reproduced in most cases, at least qualitatively, by the alternating mode visibilities, the most visible modes being symmetric with respect to the equator, i.e. their $\tilde{n}$ value is even (or $\ell=0$).

A cluster of modes is shown around $15$ c/d.  This is where the fundamental mode is expected.  We note that \citetads{2005A&A...438..633S} also found solutions for which $f_1$ was the fundamental.  In our case, it was difficult to precisely identify which of the numerically calculated modes corresponds to the fundamental mode since a large number of gravity modes (g modes) also appear in the same frequency range thus leading to multiple mode interactions and an unclear mode geometry.  Indeed, rotation causes acoustic modes, which are located in the envelope, to decrease in frequency due to the increase in equatorial radius, whereas the gravity modes, which are located more deeply in the star, are less affected.  This causes the two mode domains to overlap, in particular where the fundamental mode is located.  The modes shown in Fig.~\ref{FigAstero} have large visibilities and a relatively simple geometry in the outer portion of the star.  This interaction between g modes and the fundamental mode might explain why there are three observed pulsations in this region.  In any case, one can easily exclude the possibility that $f_1$, $f_4$, and $f_5$ are a rotational multiplet.  Indeed, the rotation rate as based on the above best-fitting models is around $2.9$ c/d which is much larger than the frequency separations between these modes.

\begin{figure*}[ht]
\centering
\includegraphics[width=\textwidth]{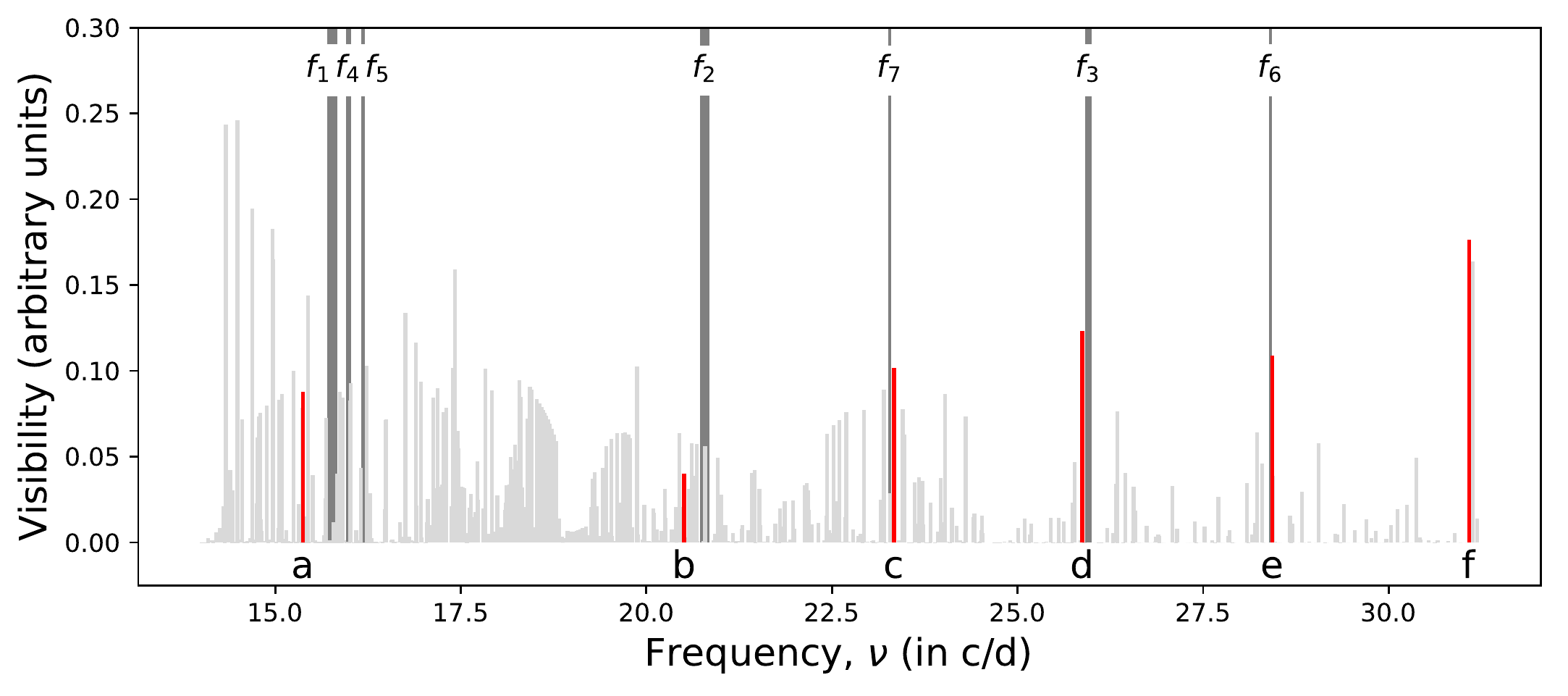}
\caption{Theoretical pulsation spectrum for the $M=1.863$\msun, $X=0.739$ model. The dark grey lines that vertically span the plot are the observed pulsations (modes $f_1$ to $f_7$). Their thicknesses are proportional to the observed amplitudes. The light grey lines of different heights are the theoretical pulsations computed from the model. The red line segments correspond to identified island (or mixed gravito-island) modes, assumed to correspond to the observed modes. Meridional cross-sections of these 6 modes, labelled with the letters `a' to `f', are shown in Fig.~\ref{FigModes}. We note that no island pulsation mode was clearly identified around 18 c/d, probably as a result of an avoided crossing.}
\label{FigAsteroFinal}
\end{figure*}

Overall, the model frequencies are somewhat too low (see left panel of Fig.~\ref{FigAstero}).  This is an indication that the mean densities of the models are too low, provided the mode identification is correct.  We therefore scaled the models using a homologous transformation to see what masses (assuming the equatorial radius is fixed as given by interferometry) would lead to a better match with the observations (specifically, we fitted the three upper pulsation frequencies).  The scaled frequencies and masses are indicated in the right panel of Fig.~\ref{FigAstero}.  As can be seen, the scaled frequencies match the observations fairly well, and the scaled masses converge towards $1.86-1.89$ \msun. For $X=0.700$, due to the correlation between $M$ and $\Xc$ (see Eq.~\ref{eq:correlation_xc}), such a mass leads to $\Xc/X$ slightly larger than unity, i.e. a core with slightly more hydrogen than the envelope. From a physical point of view, such a solution must of course be rejected. For the models at $X=0.739$, however, $\Xc/X$ remains below unity. We therefore searched for the best fitting model with $X=0.739$, thus obtaining $M=1.863$\msun. The corresponding values for $Z$ and $\Xc$ are shown in Table \ref{table:final_model}.  A full theoretical pulsation spectrum is displayed in Fig.~\ref{FigAsteroFinal} where the mode visibilities are calculated using the same normalisation as above.  As can be seen, the $(\tilde{\ell},\,m)=(0,0)$ islands modes (highlighted in red) are a good match to the observed frequencies, especially at higher frequencies.  At lower frequencies they seem to be more strongly affected by avoided crossings, thus potentially explaining the offsets between these and the observed pulsations.  At high frequencies, the island modes have the highest visilibities, whereas at low frequencies a large number of gravito-inertial modes with non-negligible visibilities are present.

\subsection{Final result}

Based on the results presented in the previous sections, we give in Table \ref{table:final_model} the parameters of our best model along with the values obtained by \citetads{2007Sci...317..342M}. These results and their uncertainties are discussed in the next section.

\begin{table}
\centering
\caption{Comparison of the fundamental parameters of Altair derived by \citetads{2007Sci...317..342M} and from our work where we use X=0.739 from \citetads{2005ASPC..336...25A}. The last column shows the parameters of the ESTER model which best reproduce all interferometric, spectroscopic and seismic data. The values in boldface are inputs of the model, and were obtained via fitting of the data, while the others are output of ESTER.}
\begin{tabular}{c | c | c} 
Parameters & \citetads{2007Sci...317..342M} & This work\\[0.5ex]
\hline
$i$ (\degr)                & 57.2 $\pm$ 1.9    & \textbf{50.65 $\pm$ 1.23}\\
$PA$ (\degr)               & 298.2 $\pm$ 0.8   & \textbf{301.13 $\pm$ 0.34}\\
$M$ (\msun)                & 1.791             & \textbf{1.86 $\pm$ 0.03}\\
$\Tpol$ (K)                & 8450 $\pm$ 140    & 8621\\
$\Teq$ (K)                 & 6860 $\pm$ 150    & 6780\\
$\Rpol$ (\rsun)            & 1.634 $\pm$ 0.011 & 1.565 $\pm$ 0.014\\
$\Req$ (\rsun)             & 2.029 $\pm$ 0.007 & 2.008 $\pm$ 0.006\\
$\veq$ (\kms)              & 285.5 $\pm$ 6     & 313\\
$\vsini$ (\kms)            & 240               & 242\\
$\Omega$ ($\Omega\sub{k}$) & 0.695 $\pm$ 0.009 & \textbf{0.744 $\pm$ 0.010}\\
$Z$                        &     --            & \textbf{0.019}\\
$[\mathrm{M}/\mathrm{H}]$  & -0.2              & \textbf{0.19}\\
$\Xc$                      &    --             & \textbf{0.71}\\
$\varepsilon$              & 0.195 $\pm$ 0.002 & 0.220 $\pm$ 0.003\\
$\beta$                    & 0.190 $\pm$ 0.012 & 0.185\\
\end{tabular}
\label{table:final_model}
\end{table}

\begin{figure*}
  \begin{minipage}[b]{.48\textwidth}
    \centering\includegraphics[width=\textwidth]{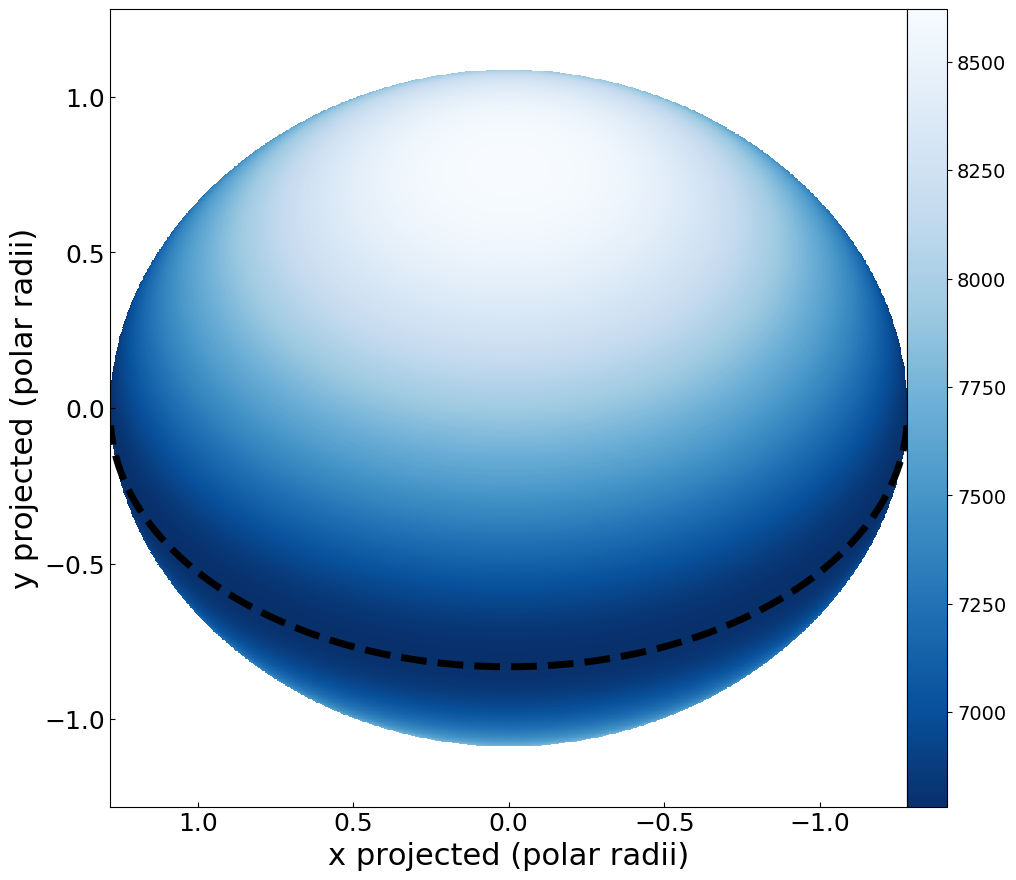}
    \caption{Surface map of the effective temperature of our best ESTER model (parameters in Table \ref{table:final_model}). The dashed line marks the equator. The values are in Kelvin.}
    \label{image:temp_map}
  \end{minipage}%
  \quad
  \begin{minipage}[b]{.48\textwidth}
    \centering\includegraphics[width=0.96\textwidth]{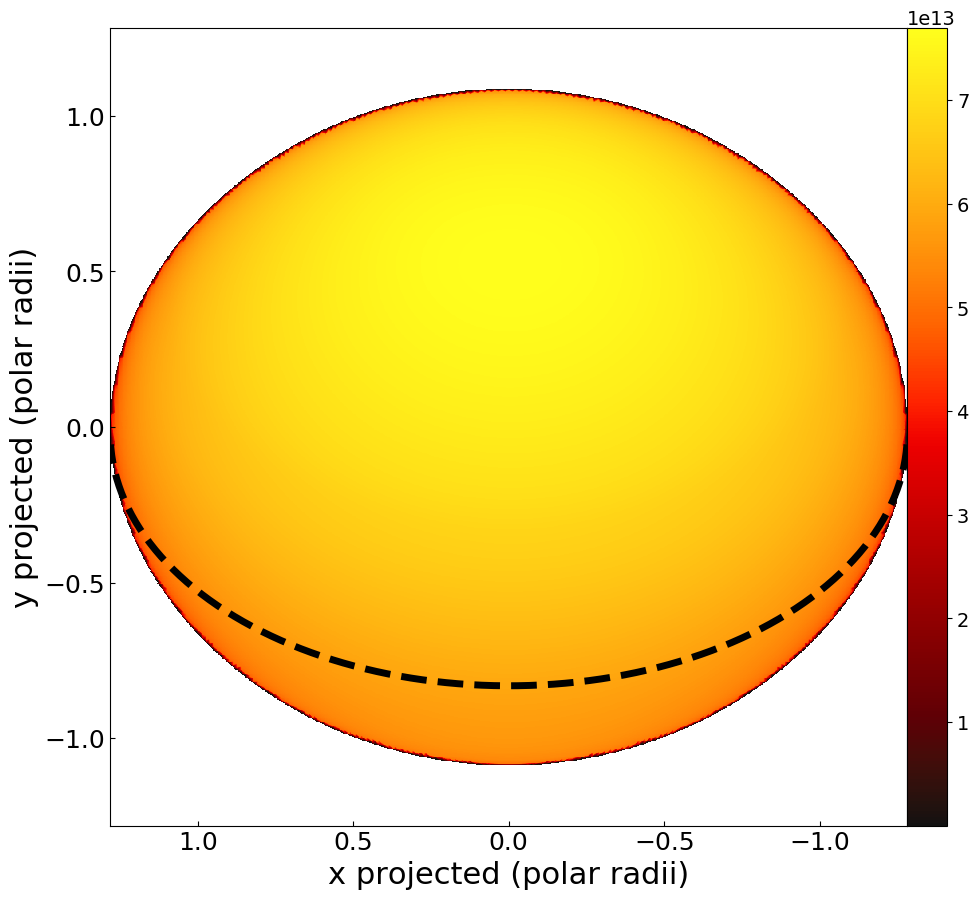}
    \caption{Monochromatic intensity map of our best ESTER model (parameters in Table \ref{table:final_model}), at 1.5 \micro\meter\ in the H band. The values are in erg\usk\reciprocal\second\usk\centi\meter\rpsquared\usk\centi\meter.}
    \label{image:int_map}
  \end{minipage}
\end{figure*}

\begin{figure*}
  \begin{minipage}[b]{.48\textwidth}
    \centering\includegraphics[width=\textwidth]{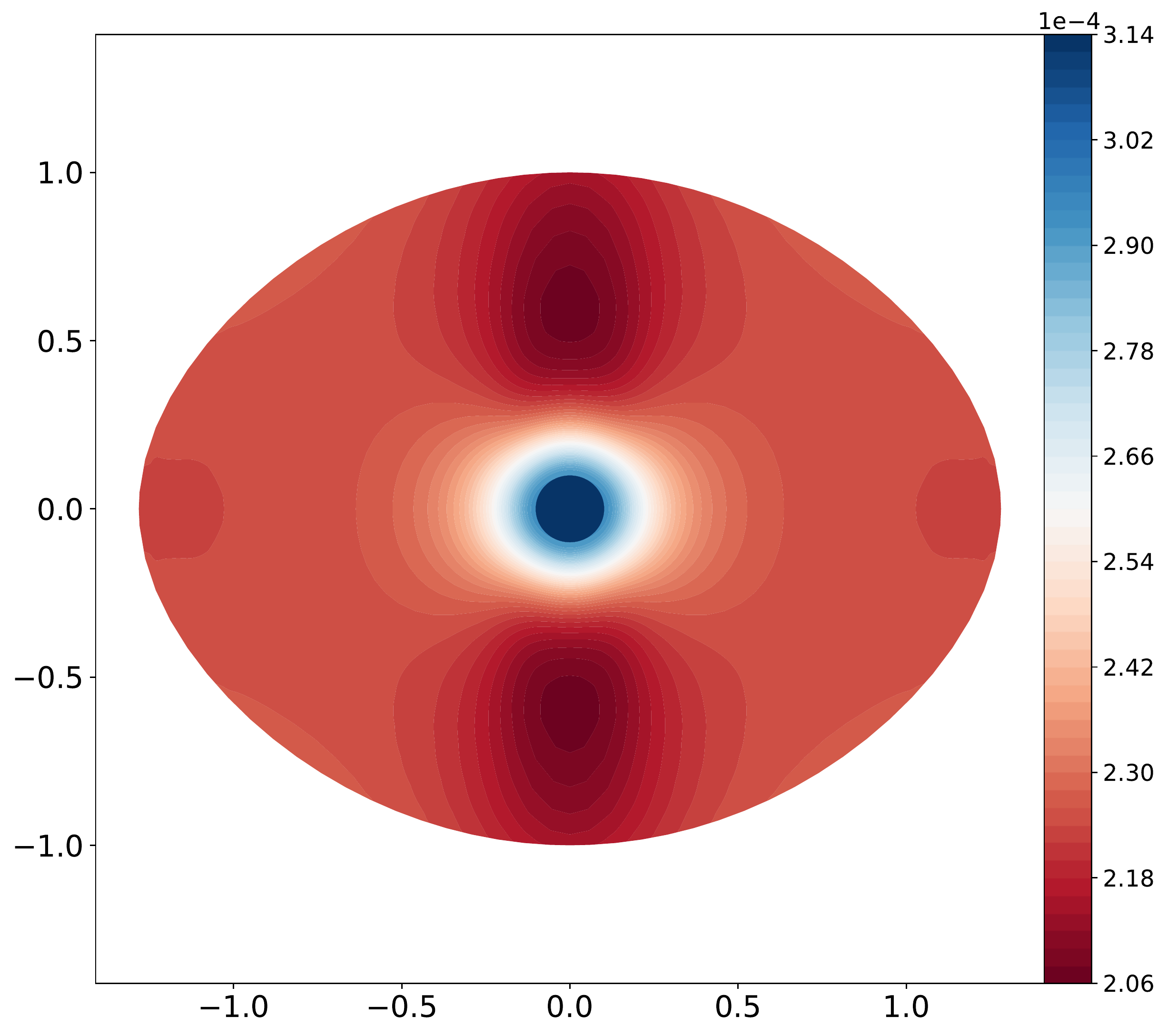}
	\caption{Meridional cut of our best ESTER model (parameters in Table \ref{table:final_model}). The colours represent the angular rotation rate.}
  \label{image:rot_rate}
  \end{minipage}%
  \quad
  \begin{minipage}[b]{.48\textwidth}
    \centering\includegraphics[width=\textwidth]{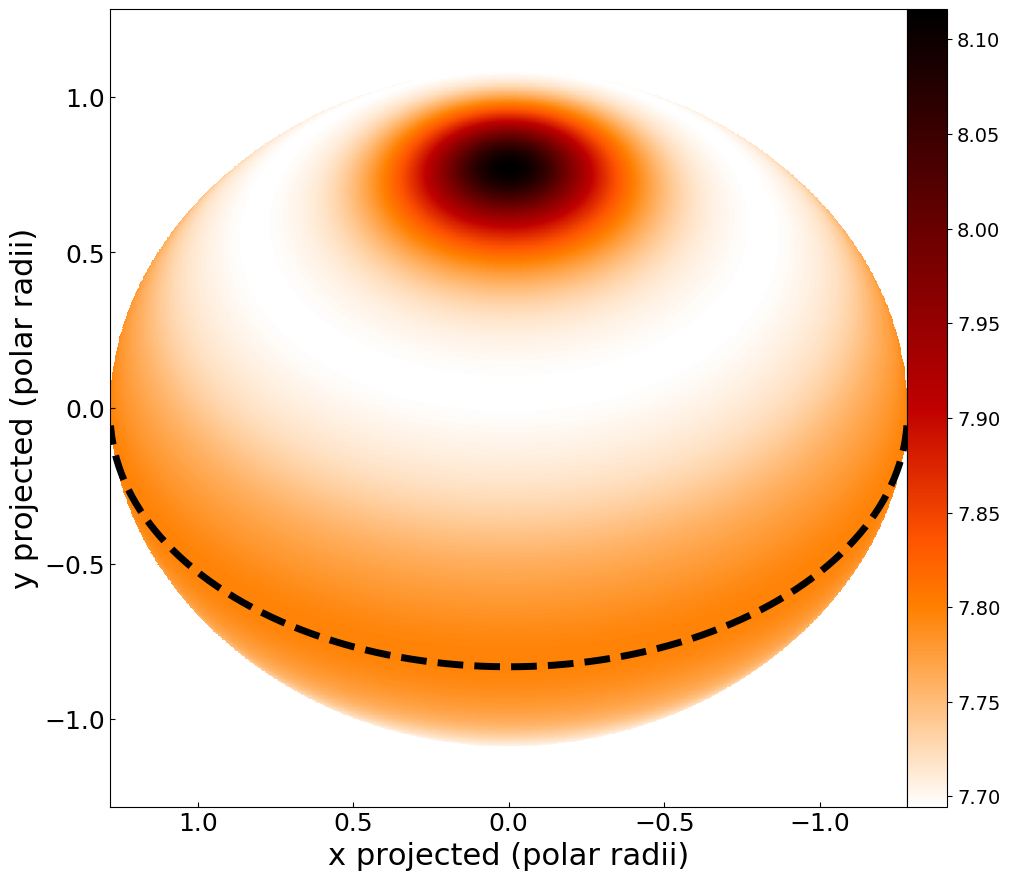}
	\caption{Surface map of the rotation rate of our best ESTER model (parameters in Table \ref{table:final_model}). The colour represents the rotation period (in hours).}
    \label{image:surf_rot}
  \end{minipage}
  
  \begin{minipage}[b]{.48\textwidth}
    \centering\includegraphics[width=0.9 \textwidth]{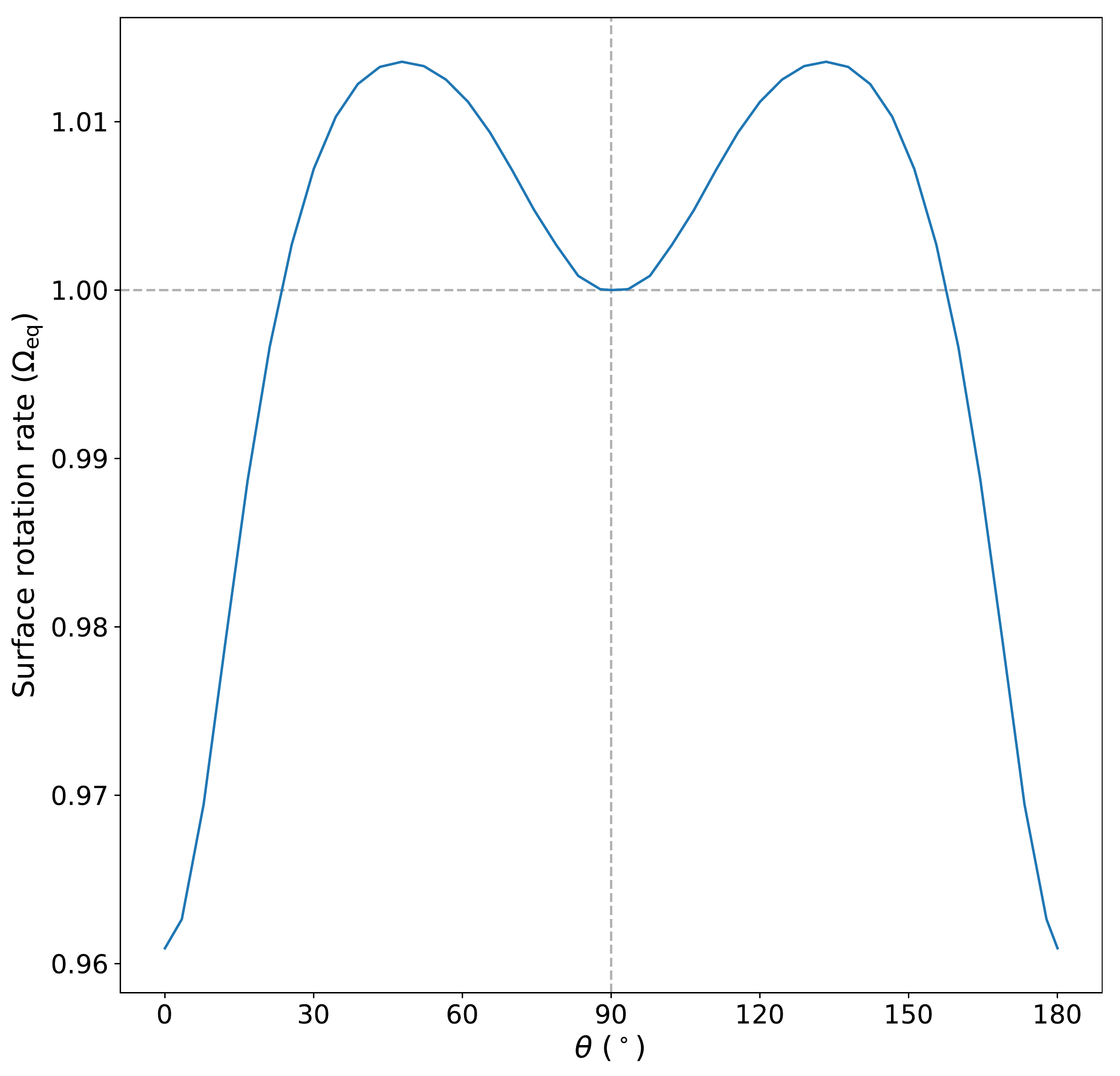}
	\caption{Surface rotation rate as a function of the co-latitude, for our best ESTER model (parameters in Table \ref{table:final_model}).}
    \label{image:omega_theta}
  \end{minipage}%
  \quad
  \begin{minipage}[b]{.48\textwidth}
    \centering\includegraphics[width=\textwidth]{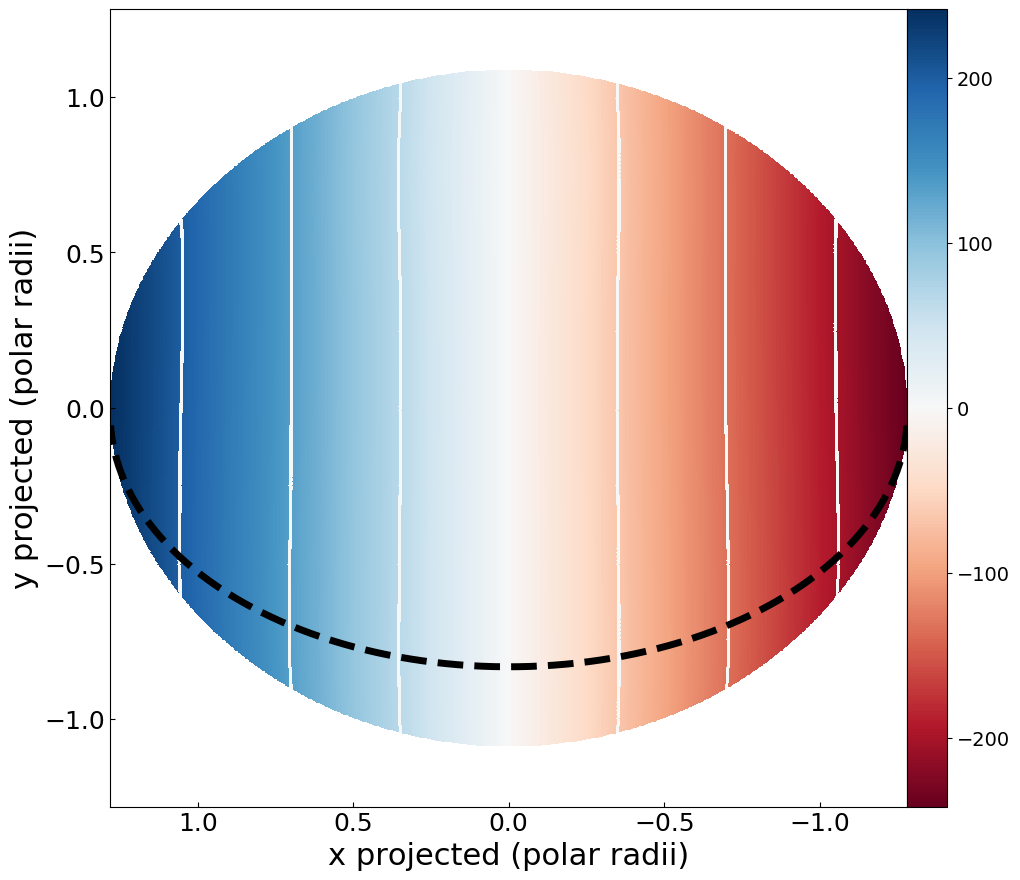}
	\caption{Surface map of the velocity of our best ESTER model (parameters in Table \ref{table:final_model}) projected onto the line of sight. The values are in \kilo\metre\usk\reciprocal\second.}
    \label{image:vproj}
  \end{minipage}
\end{figure*}

Surface maps of several relevant parameters of this model are shown in Figs. \ref{image:temp_map} to \ref{image:omega_theta}. The temperature map and the associated monochromatic intensity (in the continuum at one arbitrary wavelength in the H band) are shown in Figs. \ref{image:temp_map} and \ref{image:int_map}. Figures \ref{image:rot_rate}, \ref{image:surf_rot}, and \ref{image:omega_theta} show the internal and surface rotation profiles of the model star, while Fig. \ref{image:vproj} shows its surface velocity projected onto the line of sight. Interestingly, the slight curvature in the lines of constant projected velocity is caused by the differential rotation. The fit of the interferometric observables was already displayed in Fig. \ref{image:fit_interfero}, while Fig. \ref{image:spectrum} shows the fit to the spectrum. 

\section{Discussion} \label{section:Discussion}

\subsection{Obtained parameters}
We discuss here the parameter values obtained for our best ESTER model, shown in Table \ref{table:final_model}, and how they compare with previous estimates.

\para{Radius}
Through our analysis of interferometric data, we found for the equatorial and polar radii, $\Req$ and $\Rpol$, smaller values than both \citetads{2005A&A...442..567D} and \citetads{2007Sci...317..342M} (hereafter D05 and M07, respectively), outside of their error bars. Our polar radius especially, being 4\% smaller than that of M07, induces a higher flattening of the star in our best model ($\sim$0.22 against $\sim$0.19). This translates into a smaller apparent angular diameter in the polar direction of $\oslash\sub{p}^{\mathrm{max}} = 3.08$~mas, compared to $\sim3.30$~mas for D05 (M07 only gave an equivalent angular radius to the polar radius of their model, not the actual observed apparent angular radius in the polar direction). One can make the assumption that this effect comes directly from the difference in intensity distribution over the stellar surface compared with previous studies, which stems from the more realistic physics in our models (of great importance here is the modelling of GD). This could also explain our slightly broader range of temperatures, while our mean temperature of 7594\kelvin\ agrees well with \citetads{2003A&A...398.1121E}'s value of 7550\kelvin.

To compute the error $\Delta \Req$ on the equatorial radius, we proceeded as follows: as the distance Earth--Altair was fixed in our MCMC run, the resulting $\pm 1\sigma$ error on $R\sub{eq}$ (given by our MCMC run) corresponds in fact to an error on the angular radius $\oslash\sub{eq}$. The $\pm 0.001$\rsun\ error on $\Req$ thus corresponds to an error on $\oslash\sub{eq}$ of $\pm 0.001 \mathrm{mas}$. Coupled with the 0.3\% error on the distance given by Hipparcos (quadratically adding them), we get a relative error on the linear radius of ~0.3\%. This error is high enough to cover for the combined effects of the wavelength calibration accuracy of PIONIER ($\sim0.4$\%) and GRAVITY (a few $\sim0.01$\%). The error on the polar radius directly comes from its definition in the $\omega$-model, as a function of $\Req$ and $\Omega$.

\para{Position angle}
Also based on the interferometry, the position angle was accurately constrained to a value of 301.13 $\pm$ 0.34\degr, which is slightly above the value of 298.2\degr obtained by M07.  D05 found 298 $\pm$ 17\degr from the analysis of NPOI closure phases and squared visibilities from PTI and VINCI (Table 3, BMIRCP column), agreeing nicely with our result.

\para{$\mathbf{\Omega}$, $\mathbf{i}$, and $\mathbf{v\,sin\,i}$}
Our $\vsini$ agrees very well with that of M07, with 243 \kms\ against 240 \kms\ for them. Yet, the individual values of the inclination angle $i$ and rotation velocity $\Omega$ are heavily dependent on the brightness distribution over the surface of the star. As stated above, this distribution differs in our models from those of previous studies, leading to a lower inclination and higher rotation velocity, with the inclination still within the $2\sigma$ estimate of \citetads{2004A&A...428..199R}, that is $i>45\degr$. Errors on $i$, $\Omega$, and $PA$ are the $\pm 1\sigma$ tolerance on the MCMC model-fitting with the $\omega$-model.

\para{Mass}
Our mass determination ($M=1.863$ \msun) is significantly higher than previous determinations, with D05 citing \citetads{1990A&AS...85.1015M}'s estimate of 1.80\msun, and M07 using \citetads{2006ApJ...636.1087P}'s value of 1.791\msun. Both masses were obtained by looking for a 1D non-rotating Geneva model \citepads{1992A&AS...96..269S} which would reproduce the position of the star in the HR diagram for the former, and the corrected luminosity and polar radius estimates of the latter. Finding a $\sim 4\%$ difference in mass with their estimates, through the seismic study of a 2D stellar model which reproduces well the interferometric data, is not surprising.
The error on the mass was determined through asteroseimology. As was done in Sect. \ref{section:astero}, we computed the scaled masses needed to match the observed pulsation frequencies for models corresponding to the upper and lower boundaries for $Z$ and $\Xc$, for all $M$ and $X$ values shown in Fig. \ref{image:z_xc_intersection}. The resulting scaled masses were all between 1.84 and 1.89~\msun, that is at most 0.03 \msun\ away from the value of $M = 1.863$ \msun, which is the mass that comes out when studying the pulsation frequencies of models which verify relations \ref{eq:correlation_z} and \ref{eq:correlation_xc}, as previously stated. This is the value we adopted for the error on the mass.

\para{Temperature}
The error on the temperature is trickier. Computing the temperature of the models at $M = 1.86 \pm 0.03$~\msun\ with $Z$ and $\Xc$ following relations \ref{eq:correlation_z} and \ref{eq:correlation_xc}, for both $X=0.700$ and $0.739$, we get errors on the equatorial and polar temperature of about 40\ \kelvin\ and 55\ \kelvin\ respectively. Yet, this only accounts for the uncertainty on the mass. We can estimate an uncertainty on $Z$ and $\Xc$ by multiplying their standard deviation (computed as the square root of the diagonal elements of the covariance matrix) by the square root of the minimum of the reduced $\chi^2$, as is often done to account for the dispersion of the data. While not ideal, this method gives an idea of the uncertainties on the model parameters as a result of the dispersion of the observations with respect to the fit. This gives us a $\sim0.008$ uncertainty on $Z$, and $\sim0.185$ on $\Xc$. We may only compare the retained model (see Table \ref{table:final_model}) with the one corresponding to the lower bounds on $Z$ and $\Xc$, as the upper bound on $\Xc$ leads to $\Xc/X>1$. Doing that, we get a difference of about 640\kelvin\ at the equator, and 810\kelvin\ at the pole. This again highlights the need for an accurate determination of Altair's surface composition, as more constraints on Z and X would greatly help reduce this 9.4\% error on the temperature.

\para{$\mathbf{Z}$, $\mathbf{X}$, and $\mathbf{\Xc}$}
This considerable uncertainty on $Z$, $X$, and $\Xc$ led us not to give the error on these parameters in Table \ref{table:final_model}. 
However, our results all point towards the fact that, whatever Altair's composition is, as long as it is in the range of compositions found for stars in the vicinity of the Sun (where Altair is located), Altair is a young star, close to the Zero-Age Main Sequence (ZAMS). We used the CESAM code \citepads{2008Ap&SS.316...61M} to get an estimate of the time it would take for a 1.86\msun\, star with an initial hydrogen content $X=0.739$ to evolve from $\Xc=0.739$ to $\Xc=0.710$. The resulting age is $\sim93$ Ma. Since this age was evaluated with a 1D model taken outside its range of validity, as far as rotation is concerned, we estimate the age of Altair to be around 100 Myrs. 

\para{[M/H]}
We note that the metallicity obtained for the best ESTER model does not match that of the model atmosphere used to reproduce the observed spectrum. Indeed, the correlation between $M$ and $Z$ (see Eq.~\ref{eq:correlation_z}) leads to $Z=0.019$ for $X=0.739$. Knowing that
\begin{equation}
    \mathrm{[M/H]} = \log\left(\frac{\rm X_\sun}{\rm Z_\sun}\frac{Z}{X}\right),
\end{equation}
we get a corresponding [M/H] = 0.19 using the solar composition from \citetads{2005ASPC..336...25A}. This is lower than the metallicity used in the model atmosphere (i.e. [M/H] = 0.45).  As this mismatch is not entirely satisfactory, we carried out preliminary calculations with an atmospheric metallicity matching the bulk metallicity. This led to a slightly worse fit to the spectroscopic data and incompatible results with the interferometric constraints. Remembering that the metallicity for the atmosphere was obtained by fitting the spectroscopic data for a single absorption line of Mg\textsc{II} only, more work is needed to properly determine the atmospheric abundances of Altair. It may indeed be possible that Altair's atmosphere is more metallic than its interior, as a possible consequence of a recent accretion of metals from a residual disc or planetoids. This idea is comforted by the fact that \citetads{2017A&A...608A.113N} found an extended, weak IR excess (a few \% in K band) for Altair, which suggests the presence of a tenuous circumstellar material (possibly from a debris disc) within a few AU of the star.


\subsection{Surface convection at the equator}

\begin{figure}
\centering
  \includegraphics[width=0.48 \textwidth]{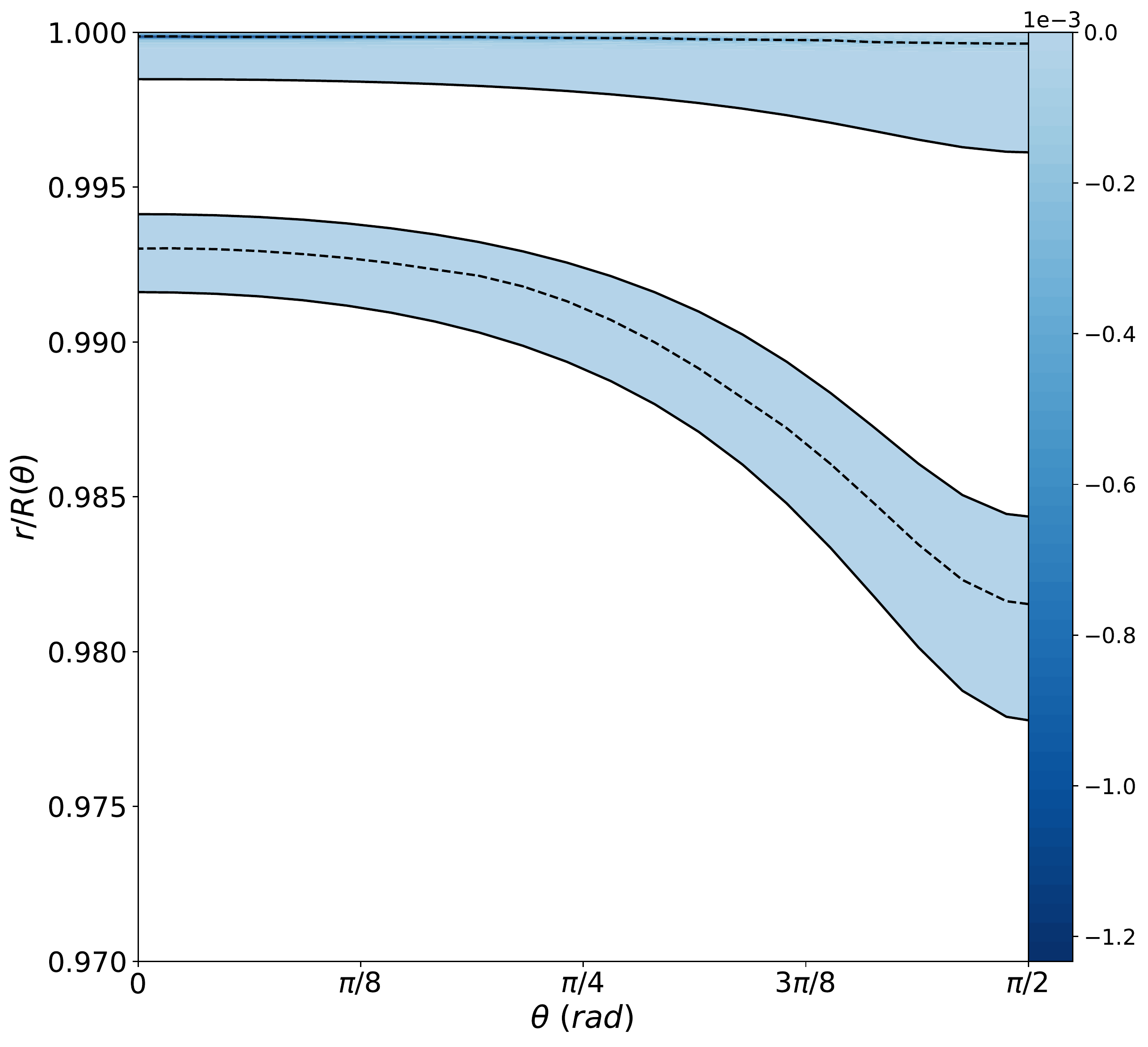}
	\caption{Squared Brunt-V\"ais\"al\"a frequency $N^2$ of the final model, as a function of the radius $r$ and colatitude $\theta$. The thermally unstable layers (regions where $N^2 < 0$), are colour-coded. These regions are bounded by solid lines, with a dashed line showing the minimum of $N^2$ in this particular region.}
\label{image:brunt_vaisala}
\end{figure}

We mentioned in Sect. \ref{section:ESTER} that ESTER models do not include surface convective layers. This is not a problem for computing the structure of intermediate-mass and massive stars, since the surface convective layers are very thin and convection carries a small fraction of the flux. However, we can still have an idea of the shapes of convective layers from the distribution of the squared Brunt-V\"ais\"al\"a frequency. In Fig.~\ref{image:brunt_vaisala} we show this distribution, which reveals two thermally unstable layers, corresponding to the hydrogen ionisation peak (reaching the surface) and helium first ionisation (below the surface). Convective layers clearly thicken at the equator, as can also be noticed in \citetads{2013A&A...552A..35E}'s model of Vega. This thickening may be related to the X-ray coronal emission of Altair, which seems to be concentrated around the equator  \citepads{2009A&A...497..511R}. Most probably, this activity is also related to Altair's observed UV emission \citepads{2005ESASP.560..903R}. Finally, surface thermal
convection likely alters the broadening of the lines, through micro and macro-turbulence, which in turn influences our fitting of Altair's spectrum.

\subsection{Resolution of the intensity maps}

The resolution of the grid used to compute the intensity maps turns out to be important. Indeed, we first decided to use only $N_\theta=25$ sectors in latitude, from pole to pole, to speed up the process. Even with such a low number of points on the visible surface of the star (876), the largest surface element is about 0.02 squared milliarcsecond, whereas the resolution achieved with the longest baseline of the GRAVITY configuration used, i.e. 130 \meter\ at $\sim2$ \micro\meter, is about 3 mas. No difference could be seen on the interferometric observables, and even though a spectrum done with a `low resolution' grid ($N_\theta=25$) was more noisy than the one done at `high resolution' ($N_\theta$ = 100), when smoothed (e.g. with a moving average), both spectra were similar (between 0.1 and 0.3\% difference). Despite all this, convergence was not reached for $\Omega$ and $i$ with $N_\theta$ = 25, even though preferred values seemed to come out of the fitting process. On the other hand, using $N_\theta$ = 100 allowed convergence towards a solution for all four parameters (see Fig. \ref{image:corner_first_four}). Unfortunately, for $\Omega$ and $i$, the solution seems to be multi-modal, as can be seen in the histograms of $\Omega$ and $i$ in Fig. \ref{image:corner_first_four}. This effect is more important at low resolution, which suggests that a higher resolution should solve the problem. We thus run the MCMC code with $N_\theta$ = 60, 80, 100, 120, obtaining 2.1, 1.6, 1.3, and 1.0 degree modulations on the inclination, respectively. This comforts us in the idea that the resolution in latitude is the main effect at play here, but as the most probable values are identical for all resolutions, we settled on a resolution of $N_\theta = 100$. This gives a high precision on all parameters (inclination included, considering the small dispersion of the peaks) while requiring reasonable numerical resources.

\subsection{Influence of the geometry of atmosphere models}

As observations became more and more accurate, limb-darkening laws derived from plane-parallel model atmospheres provided a less satisfactory fit to the observed data \citepads{2003ApJ...596.1305F, 2012MNRAS.419.1248B}. Sphericity has thus been used extensively, in recent years, in atmosphere codes, to better take into account the actual shape of a (non-rotating) star (spectra made from MARCS atmosphere models are available in both plane-parallel and spherical geometries, as is the case for PHOENIX models). Yet, when rotation comes into play, the question of the curvature radius of the model atmospheres arises. Indeed, as the rotation velocity of the star increases, so does its flattening. This gives a number of different curvature radii of the surface, as a function of latitude (the star is considered, at least in ESTER, axisymmetric). This effect is further complicated by the fact that the curvature radius is not limited to a single value at each latitude on the star, but depends on the direction one considers. Expressions of the limits on the values of the curvature radius at any point on the surface can be obtained from \citet{doi:10.1111/j.1475-1313.2006.00382.x}. Their Eq.~18 shows that the minimal and maximal values for such a radius correspond to those computed along the latitudinal and azimuthal directions. Adapting their Eq.~15 to suit ESTER's spheroidal coordinates, we find that the curvature radius in the $\theta$ direction is:
\begin{equation}
\Rc^{\theta} = -\frac{1}{\kappa(d\phi=0)} = \frac{(r^2+\rt^2)^{3/2}}{r(r-\rtt)+2\rt^2}.
\end{equation}
where $\rtt$ is the second derivative of $r$ with respect to $\theta$. Likewise, the radius of curvature in the $\phi$ direction is:
\begin{equation}
\Rc^{\phi} = -\frac{1}{\kappa(d\theta=0)} = \frac{r\sint\sqrt{r^2+\rt^2}}{r\sint - \rt\cost}.
\end{equation}

\begin{figure}
\centering
  \includegraphics[width=0.48 \textwidth]{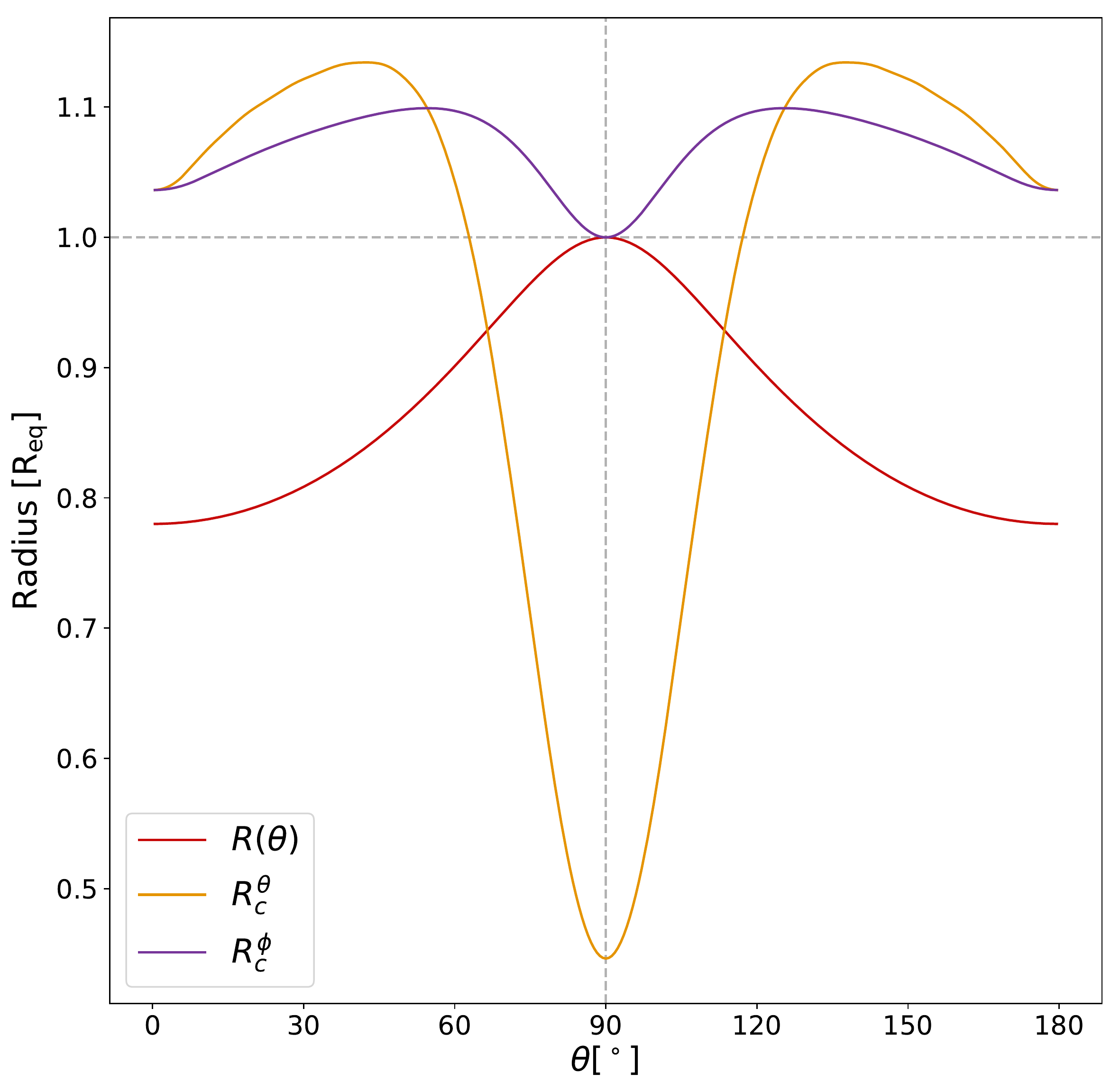}
	\caption{Curvature radii along the latitudinal ($\Rc^\theta$) and azimuthal ($\Rc^\phi$) directions, and the actual radius of the model star $r(\theta)$. The  vertical line marks the equator.}
\label{image:curvature_radius}
\end{figure}

This allows us to compute the extrema of the curvature radius from pole to pole, and compare it to the actual radius of the star (Fig. \ref{image:curvature_radius}). The figure confirms that (i) the curvature radii along the $\theta$ and $\phi$ directions are equal at the pole (as expected, since the polar axis is the axis of symmetry of the star), and (ii) the curvature radius along the azimuth is equal to the actual radius at the equator (also expected, as the equatorial plane is a plane of symmetry of the star).

If we take our ESTER model that best describes Altair, we find a curvature radius at the pole of $\Rc$ = 2.08 \rsun, and radii of $\Rc^\theta$ = 0.90 \rsun\ and $\Rc^\phi$ = 2.01 \rsun\ at the equator. As the temperature and gravity distributions at the surface of the star are not uniform, the radii of the model atmospheres associated with the different points on the grid vary, from $\Rc^{\mathrm{atm}}(\mathrm{pole}) \sim 2.0$ \rsun, to $\Rc^{\mathrm{atm}}(\mathrm{eq.}) \sim 3.0$ \rsun.
We see that, at the equator, the effective radius chosen in the atmosphere models is off the upper limit by about 1 \rsun. This and the fact itself that no point on the surface of the star (except for the poles) can be associated with a unique curvature radius make us wonder whether the need for atmospheres computed with spheroidal coordinates arises when dealing with rapid rotators.

It has been shown that the importance of the geometry of the atmosphere depends primarily on the extension of the atmosphere compared to the radius of the photosphere (see \citetads{1990MmSAI..61..765P} for example). \citetads{1991A&A...246..374B} formulated this extension $\Delta R/R\sub{photosphere}$ as being proportional to the pressure scale height, $H\sub{P}$, thus inversely proportional to the effective gravity $g$. Effectively, \citetads{2006A&A...452.1039H} determined that for $\logg > 3.0$, the estimated abundances of a number of elements are similar in both the plane-parallel and spherical geometries. \citetads{2013A&A...556A..86N} found discrepancies in limb-darkening and apparent diameters, even for stars with compact atmospheres ($\logg\geq4.0$), concluding that even in the case of `main sequence stars with large gravities and small atmospheric extensions', one should use spherical model atmospheres as they are more physically representative when trying to measure precise angular diameters and fundamental parameters of stars from optical interferometry. Yet, this effect only amounts to 1\% at maximum in the K-band, way below our uncertainty on the radius of the star. Furthermore, they also find that the difference in gravity darkening between both geometries is negligible for this kind of stars.




A star like Altair has an effective gravity at the surface $\logg > 3.0$, with our best model showing $3.8 \lesssim\logg\lesssim4.3$ from equator to pole. We should then be safe from any significant effect of the geometry of the atmospheres we used, and a spheroidal code for computing model atmospheres would not improve the accuracy of our results in a way that would justify the time and effort put into it. If surface convection and time evolution are one day successfully implemented into ESTER, then modelling evolved stars with extended atmospheres would become feasible, and this matter would have to be resolved before a work such as this one should be considered.

\section{Conclusions} \label{section:Conclusion}

We conducted a multi-technique analysis of the star Altair (HD187642) using interferometry, spectroscopy, and seismology. For the first time, this kind of analysis was performed by comparing observational data with intensity maps computed from a full two-dimensional model of the stellar interior (ESTER) and surface (atmosphere models).

PHOENIX atmosphere models were used when fitting interferometric observables, and were furthermore supplemented by Claret's 4-coefficients limb-darkening law when interpreting high resolution visible spectra from the ELODIE instrument.
For the stellar structure, the $\omega$-model (i.e. the GD model from \citetads{2011A&A...533A..43E} applied to a Roche model) first allowed us to constrain Altair's equatorial radius, position angle, rotation velocity and inclination with interferometry, while the spectrum provided a preliminary value for the metallicity of the atmosphere. Full 2D rotating stellar models from the ESTER code were subsequently used to determine the star's mass, metallicity (treated as a separate parameter from the atmospheric metallicity), and hydrogen content (both core and envelope). The correlations between these four parameters prevented convergence towards a unique solution, and the fitting of Altair's SED hardly helped. The analysis of Altair's pulsations, however, provided a solution. The observed frequencies clearly point to the upper end of the mass range and more specifically to $M=1.86$ \msun. However, due to the correlations between $M$, $Z$, $X$, and $\Xc$, such a mass would lead to $X < \Xc$ if $X = 0.700$, thus pointing towards higher values of $X$, such as $0.739$ (i.e. the solar hydrogen content based on \citetads{2005ASPC..336...25A}, as might be expected for stars in the solar neighbourhood).  Even for such a value of $X$, the value of $\Xc$ is only $4\,\%$ below that of $X$, thus indicating that Altair is young.  Even then, the solution is not unique due to the correlations between $M$, $Z$, $X$ and $\Xc$ thus pointing to the need for a full spectroscopic study, based on multiple absorption lines.  Such a study may also help resolve the difference found between the atmospheric metallicity (based on a single line) and the bulk metallicity.

This work is the first to combine such a diverse set of constraints in modelling a rapidly rotating star. It highlights the importance of using sophisticated 2D stellar models in interpreting interferometric data and is one of the very few studies which provides a plausible mode identification for acoustic pulsations in a rapidly rotating star.  Accordingly, it is also an important step in validating ESTER models from an observational point of view.  As such, it paves the way for future studies of other promising targets in a part of the HR diagram which up to now has proven challenging.

\begin{acknowledgements}
This research made use of the Jean-Marie Mariotti Center (JMMC) service \texttt{OiDB}\footnote{http://oidb.jmmc.fr}. This research made use of the SIMBAD and VIZIER databases (CDS, Strasbourg, France) and NASA's Astrophysics Data System. KB, AD, MR, and DRR acknowledge the support of the French Agence Nationale de la Recherche (ANR), under grant ESRR (ANR-16-CE31-0007-01), which made this work possible. MR and DRR also acknowledge the International Space Science Institute (ISSI) for supporting the SoFAR international team\footnote{\url{http://www.issi.unibe.ch/teams/sofar/}}.
We would also like to thank the referee for his useful insight and comments.
\end{acknowledgements}

\bibliographystyle{aa}
\bibliography{aa}

\appendix

\section{Radius of ESTER models} \label{appendix:surface_radius}

The pressure variation between the polar isobar and the photosphere (defined as where optical depth is unity) at the equator is
\begin{equation}
    \Delta P = \left(\frac{g}{\kappa}\right)_\mathrm{pole} - \left(\frac{g}{\kappa}\right)_\mathrm{eq} \simeq \frac{\Delta g}{\kappa},
\end{equation}
assuming that $\kappa_\mathrm{pole} \simeq \kappa_\mathrm{eq}$. As $\Delta g = \Omega^2 R$, we have
\begin{equation}
    \frac{\Delta R}{R} \simeq \frac{\Delta P}{\rho g R} \simeq \frac{\Omega^2}{\rho g \kappa}.
\end{equation}
For a Roche model, this means
\begin{equation}
    \frac{\Omega^2R}{g} \simeq \frac{2\varepsilon}{1 - \varepsilon},
\end{equation}
where $\varepsilon$ is the flattening coefficient. Then,
\begin{equation}
    \frac{\Delta R}{R} \simeq \frac{2\varepsilon}{1 - \varepsilon}\frac{\ell}{R}.
\end{equation}

Here $\ell$ is the mean free path of photons. For a 2~\msun\ model, $\ell = 1/(\rho \kappa) \sim 2.2\cdot10^9$ cm, and $R \sim 1.1\cdot 10^{\textrm{11}}$ cm. If the flattening is maximal ($\varepsilon = 1/3$), $\Delta R / R \sim 0.02$, if $\varepsilon = 0.2$, $\Delta R / R \sim 0.01$ and if $\varepsilon = 0.1$, then $\Delta R / R \sim 0.004$.

\section{Pulsation modes of best-fitting model} \label{appendix:modes}
\begin{figure*}[htbp]
\begin{center}
\begin{tabular}{ccc}
\includegraphics[width=0.42\textwidth]{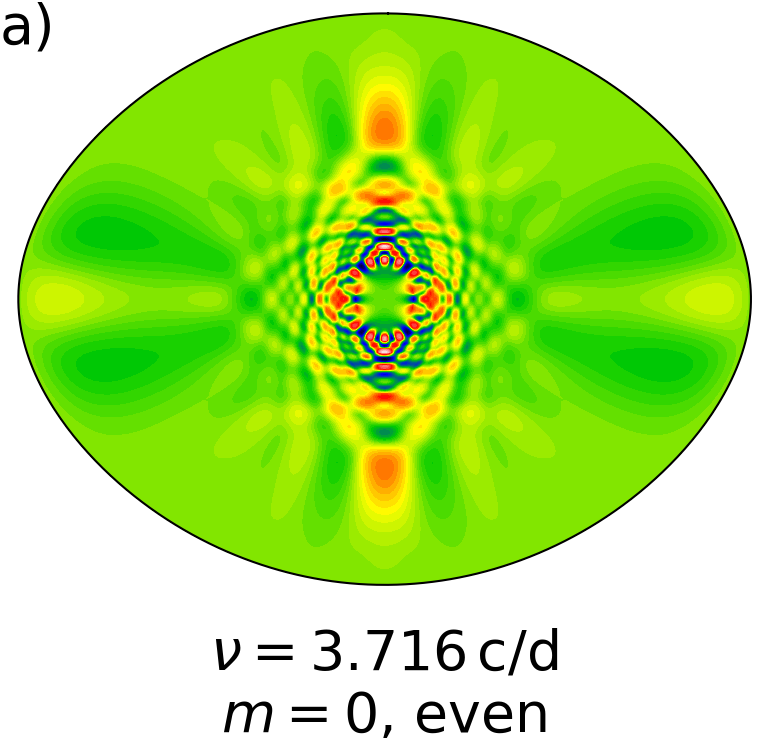} & ~&
\includegraphics[width=0.42\textwidth]{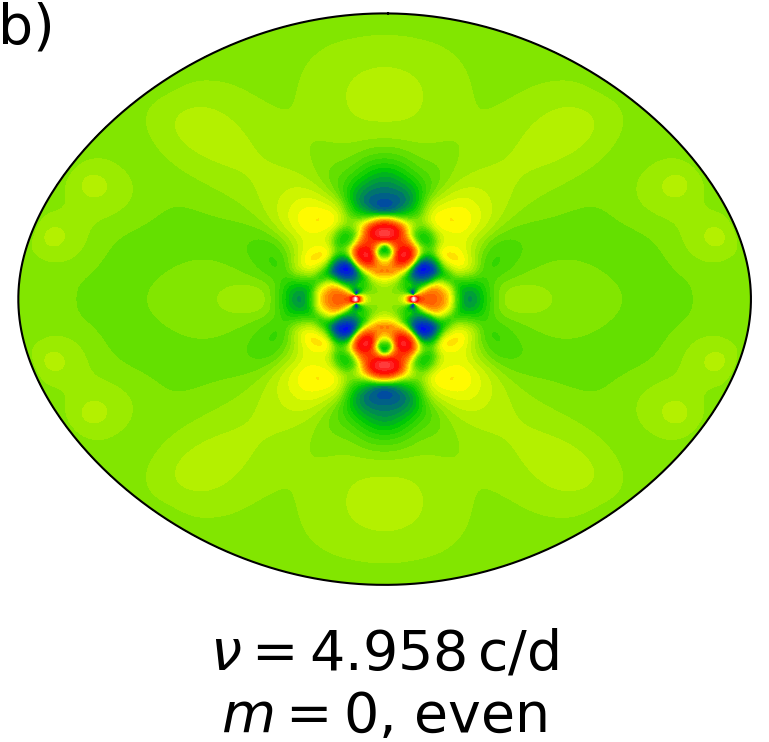} \\
\includegraphics[width=0.42\textwidth]{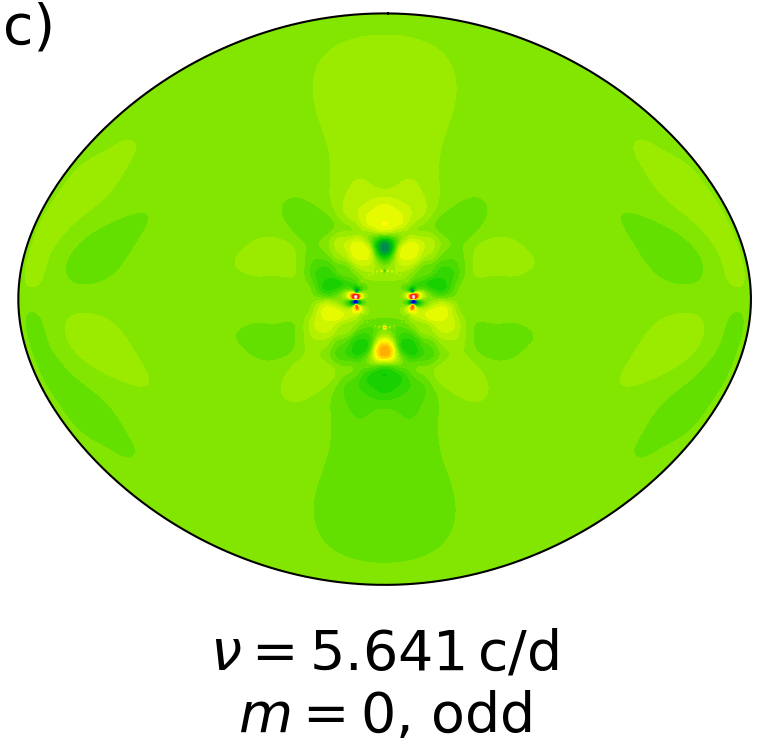} & &
\includegraphics[width=0.42\textwidth]{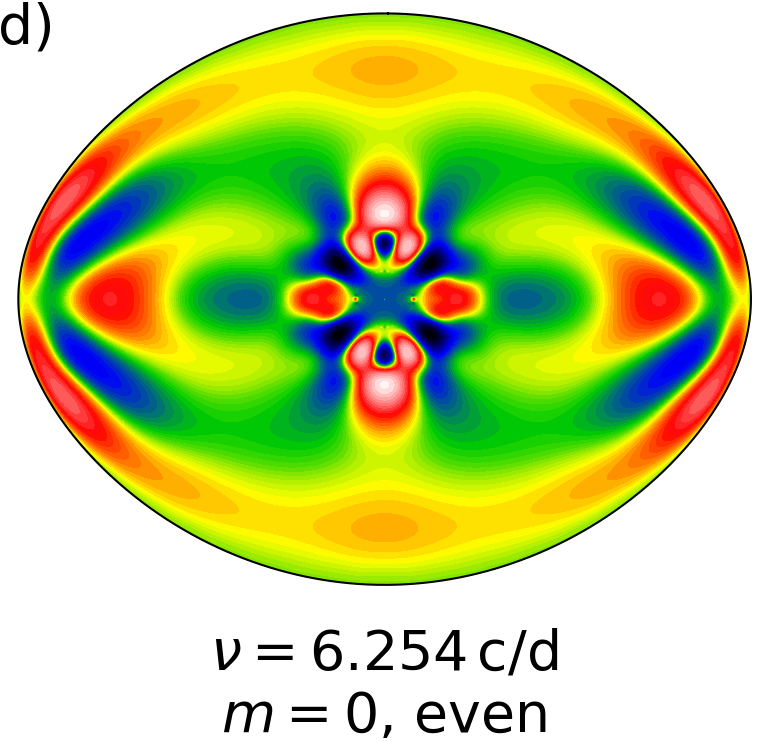} \\
\includegraphics[width=0.42\textwidth]{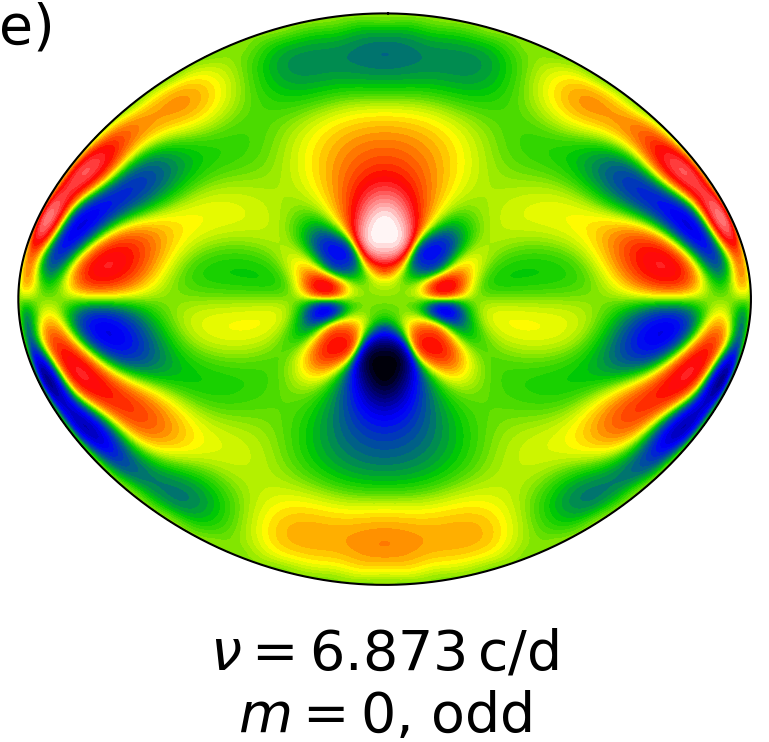} & &
\includegraphics[width=0.42\textwidth]{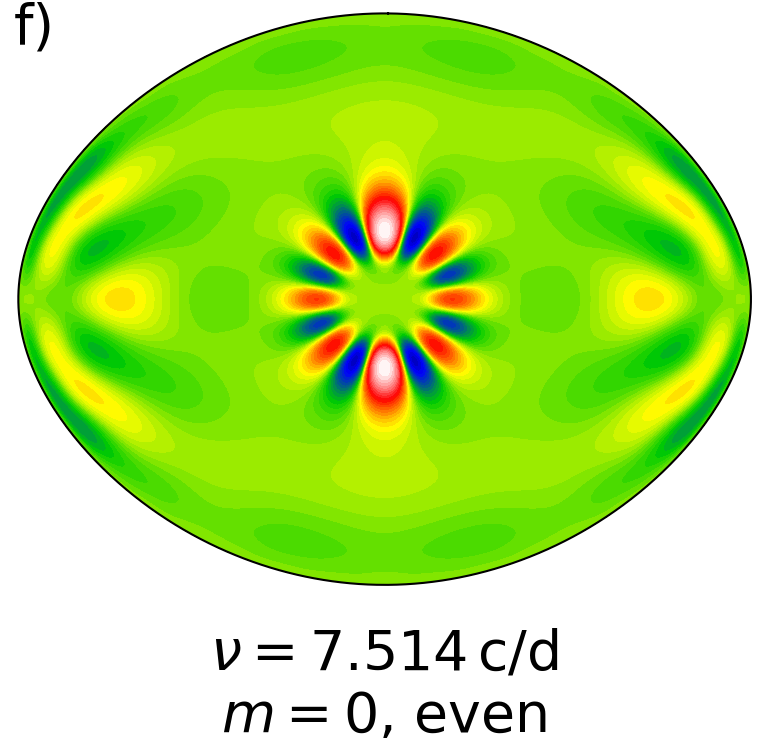}
\end{tabular}
\caption{Meridional cross-sections of island (or mixed gravito-island) modes in best-fitting model ($M=1.863$\msun, $X=0.739$).  The Lagrangian pressure perturbation, normalised by the square root of equilibrium pressure, is shown.  A pseudo-logarithmic colour scale is used to bring out faint details.}
\label{FigModes}
\end{center}
\end{figure*}

Figure~\ref{FigModes} shows the meridional cross-sections of six island modes from the best-fitting model.  A pseudo-logarithmic colour scale\footnote{Specifically, we normalise the amplitude such that its maximum absolute value is $10$, then apply the function $f(x) = \mathrm{sgn}(x) \ln\left(1+|x|\right)/\ln(11)$ prior to visualisation.}  is used to bring out faint details.  As can be seen, all of these modes are mixed with gravity modes, or even a rosette mode for the first one.

\end{document}